\documentclass[sn-mathphys]{sn-jnl}

\jyear{2021}%
\theoremstyle{thmstyleone}%
\theoremstyle{thmstyletwo}%
\makeatletter
\newenvironment{figurehere}
{\def\@captype{figure}}
{}
\makeatletter

\theoremstyle{thmstylethree}%
\usepackage{graphicx}
\usepackage{longtable}
\usepackage{xspace}
\usepackage[capitalise]{cleveref}

\newcommand{\vn}{{\mathbf{n}}}

\newcommand{\vv}{{\mathbf{v}}}

\newcommand{\vx}{{\mathbf{x}}}

\newcommand{\vz}{{\mathbf{z}}}

\newcommand{\vTheta}{{\boldsymbol{\Theta}}}

\newcommand{\vC}{{\mathbf{C}}}

\newcommand{\vV}{{\mathbf{V}}}

\newcommand{\EE}{\mathbb{E}}

\newcommand{\KL}{\mathrm{KL}}

\usepackage{makecell}

\begin{document}

\title[Article Title]{Addressing preferred orientation in single-particle cryo-EM through AI-generated auxiliary particles}


\author[1]{\fnm{Hui} \sur{Zhang}}
\equalcont{These authors contributed equally to this work.}

\author[2]{\fnm{Dihan} \sur{Zheng}}
\equalcont{These authors contributed equally to this work.}

\author[6,7,8]{\fnm{Qiurong} \sur{Wu}}
\author[5,6,7,8,9]{\fnm{Nieng} \sur{Yan}}

\author*[2,3]{\fnm{Zuoqiang} \sur{Shi}}\email{zqshi@tsinghua.edu.cn}
\author*[5,6]{\fnm{Mingxu} \sur{Hu}}\email{humingxu@smart.org.cn}
\author*[2,3,4]{\fnm{Chenglong} \sur{Bao}}\email{clbao@mail.tsinghua.edu.cn}
\affil[1]{\orgdiv{Qiuzhen College}, \orgname{Tsinghua University}, \orgaddress{\city{Beijing}, \country{China}}}
\affil[2]{\orgdiv{Yau Mathematical Sciences Center}, \orgname{Tsinghua University}, \orgaddress{\city{Beijing}, \country{China}}}
\affil[3]{\orgname{Yanqi Lake Beijing Institute of Mathematical Sciences and Applications}, \orgaddress{\city{Beijing}, \country{China}}}
\affil[4]{\orgname{State Key Laboratory of Membrane Biology}, \orgdiv{School of Life Sciences}, \orgdiv{Tsinghua University}, \orgaddress{\city{Beijing}, \country{China}}}
\affil[5]{\orgname{Shenzhen Medical Academy of Research and Translation (SMART)}, \orgaddress{\city{Shenzhen}, \country{China}}}
\affil[6]{\orgname{Beijing Frontier Research Center for Biological Structure (Tsinghua University)}, \orgaddress{\city{Beijing}, \country{China}}}
\affil[7]{\orgdiv{Tsinghua-Peking Joint Center for Life Sciences}, \orgdiv{Tsinghua University}, \orgaddress{\city{Beijing}, \country{China}}}
\affil[8]{\orgdiv{School of Life Sciences}, \orgdiv{Tsinghua University}, \orgaddress{\city{Beijing}, \country{China}}}
\affil[9]{\orgdiv{Department of Molecular Biology}, \orgdiv{Princeton University}, \orgaddress{\city{Princeton, NJ}, \country{USA}}}


\abstract{The single-particle cryo-EM field faces the persistent challenge of preferred orientation, lacking general computational solutions. We introduce cryoPROS, an AI-based approach designed to address the above issue. By generating the auxiliary particles with a conditional deep generative model, cryoPROS addresses the intrinsic bias in orientation estimation for the observed particles. We effectively employed cryoPROS in the cryo-EM single particle analysis of the hemagglutinin trimer, showing the ability to restore the near-atomic resolution structure on non-tilt data. Moreover, the enhanced version named cryoPROS-MP significantly improves the resolution of the membrane protein Na\textsubscript{X} using the no-tilted data that contains the effects of micelles. Compared to the classical approaches, cryoPROS does not need special experimental or image acquisition techniques, providing a purely computational yet effective solution for the preferred orientation problem. Finally, we conduct extensive experiments that establish the low risk of model bias and the high robustness of cryoPROS.}


\maketitle

\section{Introduction}\label{sec1}

 Recent advancements in cryo-EM hardware and image processing software have ushered in a transformative era in structure determination, establishing it as the prevailing method in the structural biology domain. Notably, Jacques Dubochet's immersion freezing workflow has become a foundational aspect of cryo-EM technology, demonstrating broad applicability across a range of biological specimens~\cite{DUBOCHET1985,dubochet1988}. Yet, challenges remain: factors such as ice quality on the grid and particle distribution within the ice can significantly impact the results. Consequently, preparing high-quality samples continues to be a critical requirement for successful 3D structure determination through single-particle cryo-EM~\cite{passmore_2016}. 
 
A main challenge during the stage of frozen samples that impedes structural analysis is the issue of preferred orientation~\cite{barth_approximation_1989, boisset_overabundant_1998, glaeser_how_2016, tan_addressing_2017, glaeser_opinion_2017, lyumkis_challenges_2019, glaeser_how_2019}. In an ideal scenario, biomacromolecules of interest should exhibit uniformly random orientations within the vitreous amorphous ice. However, it is commonly observed that samples tend to adopt a specific, dominant orientation due to interactions at the air-water or the support-water interface~\cite{noble_routine_2018, drulyte_approaches_2018}. This phenomenon can significantly degrade the resolution and quality of the resulting map, often leading to misleading density maps known as preferred orientation artifacts. Addressing this preferred orientation issue is widely recognized as an important step in refining cryo-EM as a more universal method for structural analysis~\cite{glaeser_how_2016, glaeser_how_2019, carragher_current_2019, cianfrocco_what_2020, li_effect_2021, saibil2022cryo, liu_better_2023}, making it the go-to technique for high-throughput biomolecular research and drug screening~\cite{li2020high, lovestam2022high, shen20182017}.

Over the years, numerous attempts have been made to confront the issue of preferred specimen orientation in cryo-EM, with most efforts centered around grid preparation and data collection. Techniques such as the use of detergents~\cite{lyumkis_cryo-em_2013, chowdhury_structural_2015, li_effect_2021, frotscher_fluorinated_2015}, ice thickening~\cite{huntington_thicker_2022}, and specific biomolecule modifications~\cite{bromberg_his-tag_2022} have been explored. While these methods showed promising results on specific proteins, they were often encumbered by extensive condition screening, added complexities, and time-intensive protocols without consistent success. An alternative proposition suggested using hydrophilic or selectively capturing modified graphene as a support to standardize the particle pose distribution~\cite{xu_structural_2021, wu_application_2021}. However, the preparation and modification of graphene proved labor-intensive and achieved limited success in fully resolving the orientation problem. In addition, compensatory techniques during data collection, particularly tilting strategies~\cite{tan_addressing_2017}, were introduced. These methods bypassed the challenges of sample preparation but introduced other drawbacks, such as decreased image acquisition efficiency, escalated beam-induced movement during tilting, the imperative of precise defocus gradient estimation, and increased ice layer thickness due to geometric constraints~\cite{drulyte_approaches_2018}. For small molecules, aggregation complications also arose, somewhat limiting the effectiveness of the tilting methodology. Despite these difficulties, tilting is currently one of the most effective solutions to tackle preferred orientation in cryo-EM studies. Given the aforementioned limitations in addressing the preferred orientation during the wet-lab phase, designing the computational approach becomes an attractive alternative, which motivates our work. 

We first conduct extensive experiments demonstrating that the main challenge in addressing the preferred orientation issue lies in the misalignment of particle images during the refinement process due to unevenly distributed signals from different views. This inaccurate orientation estimation leads to poor reconstruction quality, particularly impacting the axial resolution. In contrast, the missing wedge issue in computational tomography (CT) lacks projected data from certain view angles; however, the corresponding orientation information for the observed particles is always accurately known. Therefore, equating the missing wedge issue in CT with the reconstruction artifacts observed in cryo-EM single particles exhibiting preferred orientation is a misconception. Such conflation obscures our accurate understanding of the problem. Sorzano \textsl{et al.} have posited this claim~\cite{sorzano_clustering_2010, sorzano_algorithmic_2021, sorzano_bias_2022}, and we further validate its soundness.

Motivated by the above observation, we introduce cryoPROS (\textbf{PR}eferred \textbf{O}rientation \textbf{S}olver), an AI-based computational approach designed to address preferred orientation challenges in cryo-EM. Using a conditional variational auto-encoder (CVAE) model, cryoPROS generates high-quality auxiliary particles with diverse orientations, significantly improving the alignment accuracy of the observed particles. As a pure computational framework, cryoPROS does not require specialized sample preparations, specific data acquisition methods, or further computational refinement, substantially reducing experimental complexities. Our experimental results demonstrate that cryoPROS can obtain the near-atomic resolution structure using the highly preferred oriented dataset of hemagglutinin (HA) trimer (EMPIAR-10096). Furthermore, for the samples that are embedded in detergent micelles or lipid nanodiscs, we propose a variant of cryoPROS named cryoPROS-MP. This enhanced version has demonstrated a significant resolution improvement over conventional methods using the non-tilted data of the membrane protein $\text{Na}_{\text{X}}$. Finally, we test various configurations that confirm the low model bias and the high consistency of cryoPROS. This makes cryoPROS an effective and reliable tool, ideally supplementing traditional structural determination methodologies and extending its range of applications. 

\section{Results}\label{sec2}

We next present a comprehensive analysis of the preferred orientation problem, followed by introducing our novel algorithm, cryoPROS, designed to address this challenge. We illustrate the practical implementation of cryoPROS on multiple datasets and examine its characteristics and performance in-depth.

\subsection{Orientation estimation bias: the main issue for structure determination from data with preferential orientation.}\label{sec:orientation_restoration_bias}
\begin{figure}[p]
\begin{center}
\includegraphics[width=1\textwidth]{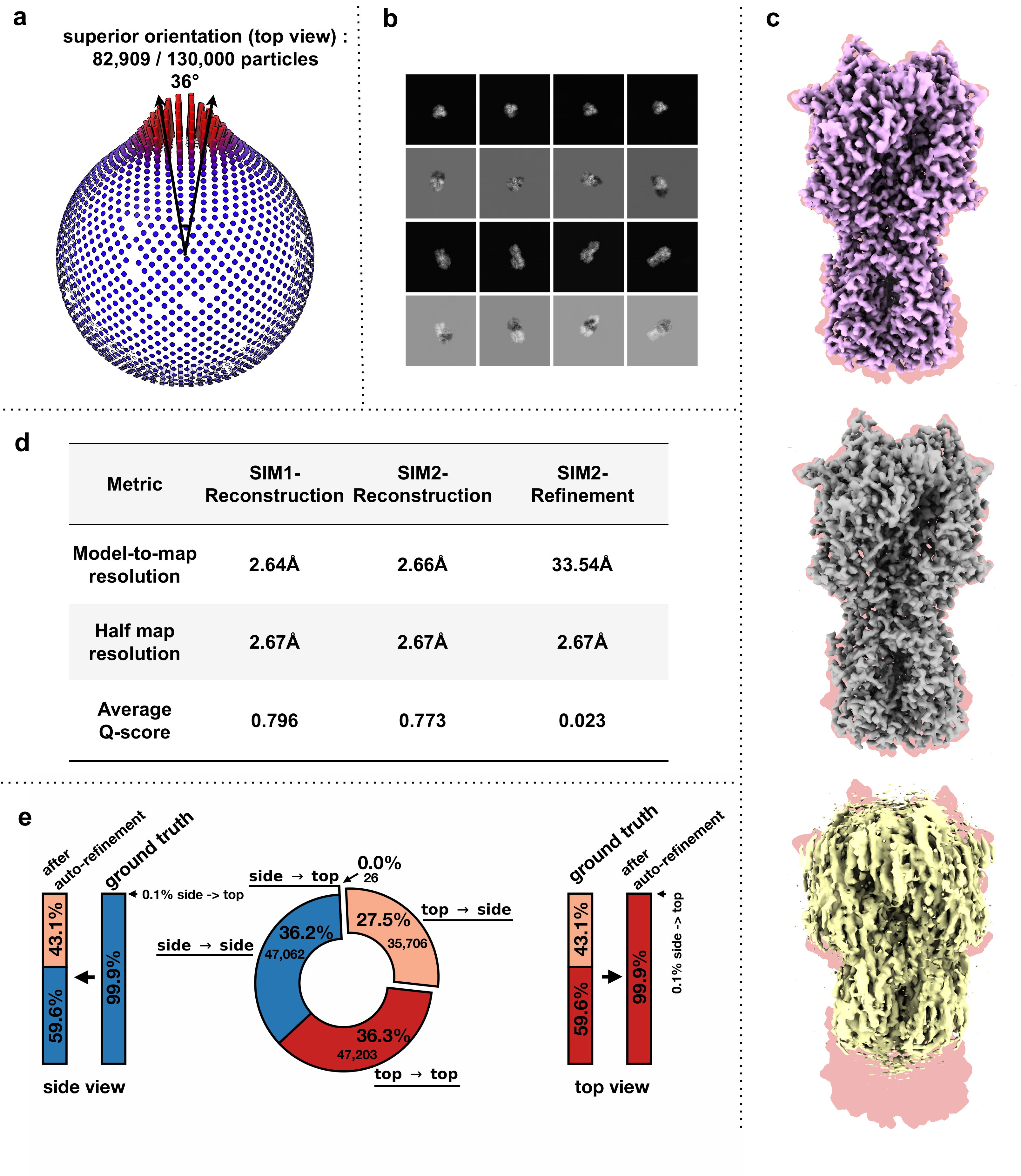}
\caption{\textbf{Effect of bias in orientation estimation on determining structures from preferentially oriented data.} \textbf{a}, Pose distribution of SIM2 dataset, emphasizing the superior orientation defined as being within $36^{\circ}$ of the Z-axis. The proportion of this superior orientation is indicated. \textbf{b}, From the SIM2-Refinement dataset, eight particles were chosen, represented by eight pairs of top-bottom images, resulting in a total of 16 images arranged in a $4 \times 4$ grid. For each pair, the top image shows the ground-truth projection (without added noise) of the HA trimer, while the bottom image represents the difference between the projection (with its orientation estimated by CryoSPARC’s autorefine) and the ground-truth projection. Among these eight particles, the ground truth orientations for half are from the superior orientation (top view), while the other half is from the inferior orientation (side view). \textbf{c}, Density maps are presented from top to bottom in the following order: SIM1-Reconstruction (in violet), SIM2-Reconstruction (in gray), and SIM2-Refinement (in yellow). \textbf{d}, Quantitative comparison of indicators for the density maps displayed in \textbf{c}. \textbf{e}, In SIM2-Refinement, CryoSPARC’s autorefine incorrectly estimated $59.6\%$ of the particles from a superior (top view) to an inferior orientation (side view), as shown on the right. These misassignments account for $35,706$ particles or $27.5\%$ of the total, overwhelming the actual side view, as highlighted in the middle. In SIM-Refinement, misassigned top views made up $43.1\%$ of the side view orientation, as illustrated on the left.}
\label{fig: orientation estimation bias}
\end{center}
\end{figure}

We conducted a series of simulated tests to identify the primary computational challenge associated with the preferred orientation problem. To do this, we synthesized two datasets, called SIM1 and SIM2, by projecting the density map of the HA trimer. The density map was derived from the atomic model (PDB ID: 3WHE) and the projection process was implemented using the {\texttt relion\_project} module within Relion. We added Gaussian noise to the particles, with a zero mean and a standard deviation of 60. Both SIM1 and SIM2 contain 130,000 particles, and the average signal-to-noise ratio (SNR) is -15.63db. In SIM1, the orientation of particles was sampled from a uniform distribution in $\mathrm{SO}(3)$. In contrast, the particles in SIM2 exhibit a preferred orientation, with 82,909 particles oriented near the Z-axis, as shown in (\cref{fig: orientation estimation bias}a). We reported the results under the following three procedures: 
\begin{enumerate}
    \item SIM1-Reconstruction: reconstruction of SIM1 using true orientations;
    \item SIM2-Reconstruction: reconstruction of SIM2 using true orientations;
    \item SIM2-Refinement: reconstruction of SIM2 using orientations estimated by the autorefine module in CryoSPRC.
\end{enumerate}
The reconstructed density maps are presented in (\cref{fig: orientation estimation bias}c). The density maps from SIM1-Reconstruction and SIM2-Reconstruction are similar, exhibiting comparable quantitative performance in terms of FSC-based metrics and Q-score (\cref{fig: orientation estimation bias}d). However, the density map produced by SIM2-Refinement shows significant deterioration, particularly in the loss of axial density due to the preferred orientation (\cref{fig: orientation estimation bias}c and \cref{fig: orientation estimation bias}d).
This degradation appears to be induced by the misalignment (\cref{fig: orientation estimation bias}e) during the refinement process. In SIM2-Refinement, approximately 43.1\%, $35,706$ out of $82,709$, of the particles with a superior orientation (top view) were incorrectly estimated as having an inferior orientation (side view) by CryoSPARC’s autorefine (\cref{fig: orientation estimation bias}e, right). As a result, these incorrectly assigned particles of the top view, constituting around 43.1\% of the side view orientation (\cref{fig: orientation estimation bias}e, left) in the reconstruction stage, significantly impacted the resolution.

The above observations motivated us to re-evaluate the challenges posed by the data with preferred orientations. The conventional approach to single-particle cryo-EM structure determination is an iterative process, starting with particle alignment through projection matching and followed by reconstruction. These steps generate estimated orientations and density maps, respectively. However, this method encounters considerable difficulties when addressing preferentially-oriented datasets. In the reconstruction phase, uneven coverage of the Fourier space results in reconstructions with anisotropic resolution. Furthermore, when a density map from the previous iteration is employed as a refinement reference, it inevitably leads to misalignment. This misalignment introduces orientation estimation errors, further reducing the quality of the reconstruction. As this iterative process continues, orientation estimation errors accumulate, leading to an increasing bias. This accumulating bias either progressively deteriorates the density or results in entirely erroneous densities. When overfitting occurs, it accentuates this decline, further undermining the reconstruction quality. In summary, the challenges of preferred orientation in single-particle cryo-EM primarily stem from orientation estimation bias. Additionally, we expanded our investigation to include real data, as detailed in \cref{sec:TRPA1}. These findings further supported the above analysis related to datasets with preferred orientations and inspired us to explore computational approaches that can accurately determine the poses of particles.

\subsection{The cryoPROS method.}\label{sec:architecture}

\begin{figure}[htp]
\begin{center}
\includegraphics[width=1\textwidth]{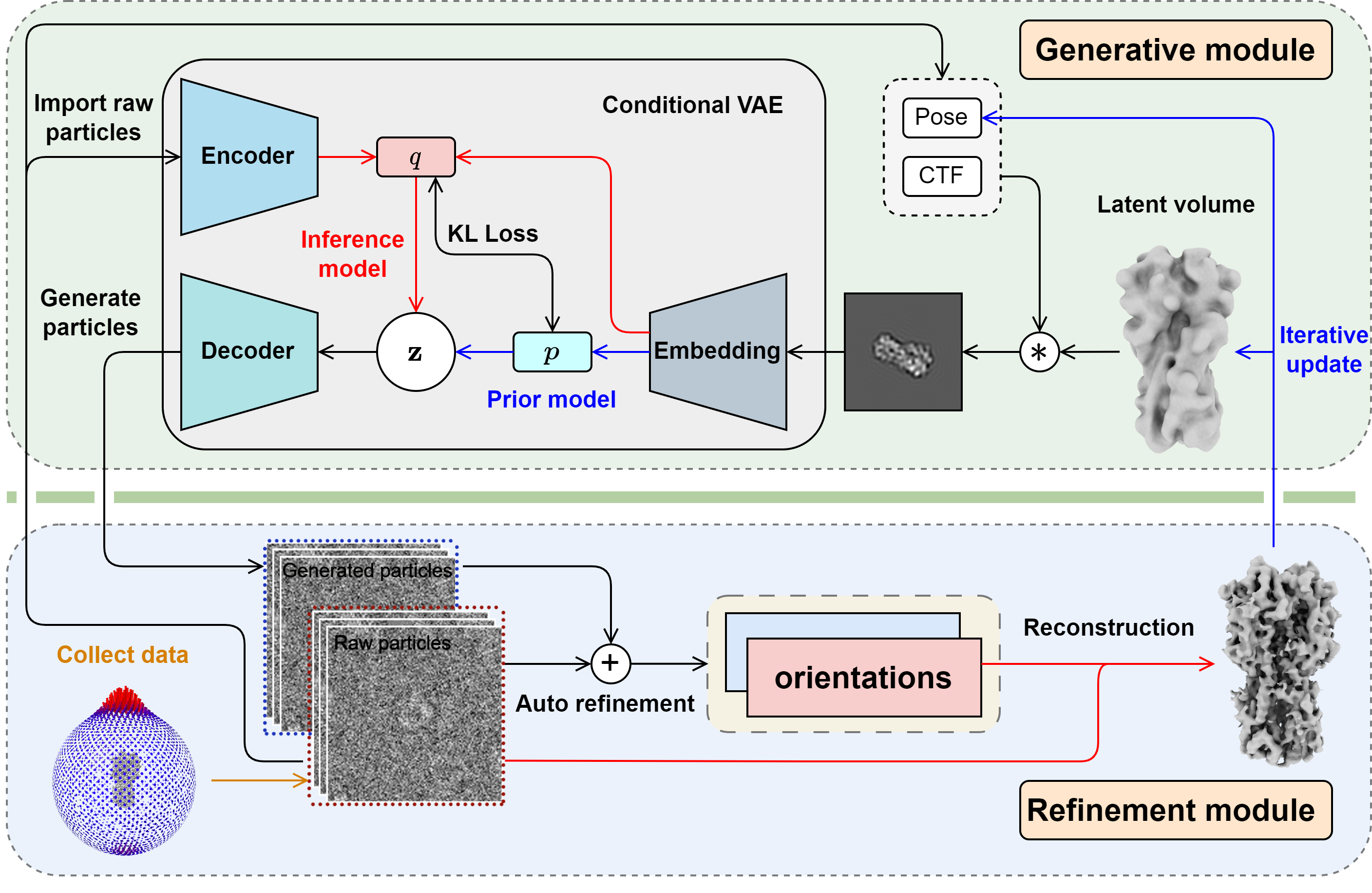}
\caption{\textbf{The cryoPROS method for single-particle cryo-EM preferred orientation problem.} The overall protocol of cryoPROS contains generative and refinement modules, applied iteratively. The generative module employs a conditional variational autoencoder (CVAE) framework, incorporating a latent volume in the latent space and integrating a physical imaging process. It generates projection particles with predetermined imaging parameters after training. The refinement module combines raw particles with generated particles, utilizing cryo-EM single-particle analysis software for \textsl{ab initio} orientation estimation. The reconstruction employed only the raw particles, omitting the generated ones, and used the updated orientations to produce a more precise 3D density map.}
\label{fig:method}
\end{center}
\end{figure}

We introduce cryoPROS, an AI-based method designed for accurate pose estimation in preferentially oriented data. Specifically, cryoPROS consists of two main modules: the generative module and the refinement module, as illustrated in~\cref{fig:method}. The generative module leverages a Conditional Variational Autoencoder (CVAE) framework~\cite{kingma2013auto,NIPS2015_8d55a249}, taking as its inputs a highly preferentially oriented raw particle stack, estimated imaging parameters, and a 3D latent volume. By minimizing the conditional evidence lower bound, we establish a self-supervised loss that eliminates the need for an additional supervised dataset (see Methods). After training, the network can generate auxiliary particles with uniformly distributed orientations. The refinement module combines the raw particles with the generated particles and employs cryo-EM single-particle analysis software (e.g., Relion, cryoSPARC, or cisTEM) to perform ab initio reconstruction and pose estimations. It is worth mentioning that the generated particles are only used for estimating poses, not for reconstructing the density map.

The CVAE model in cryoPROS consists of two components: the conditional encoding stage and the decoding stage. In the conditional encoding stage, we integrate the encoding features of both the raw particle and the latent particle using given imaging parameters (CTFs, poses) and a latent volume to form an inference model $q$ for the latent variable $\vz$. Sampling from the inference model $q$, the decoding network maps the latent variable to a synthesized particle, thereby generating a reconstruction loss in relation to the input raw particle. Moreover, we directly encode the latent particle to obtain a prior model $p$ and impose a loss based on the Kullback-Leibler (KL) divergence between the inference model $q$ and the prior model $p$. After training, we can generate auxiliary particles by decoding samples from the prior model $p$, inputting the inferior orientations and CTF parameters. To enhance its expressive ability, we designed a hierarchical VAE structure~\cite{NEURIPS2020_e3b21256,child2021very} (see Methods and Supplementary Section 6).

Challenges arise when target proteins are situated in micellar environments not present in the homologous protein or associated with different micellar contexts. In such cases, reconstructing the relevant micellar environment for the low-pass filtered homologous protein, used as the initial latent volume, becomes crucial. Therefore, we propose cryoPROS-MP, which is customized for membrane proteins in detergent micelles or lipid nanodiscs. Compared to cryoPROS, cryoPROS-MP includes an additional procedure for reconstructing the micellar information (see Methods).

\subsection{Validation with several preferential oriented datasets.}\label{sec:DATA}

We conducted validation of cryoPROS on SIM2 and three experimental datasets: the PO-subset of TRPA1, obtained by removing side view particles from the TRPA1 dataset (EMPIAR-10024); the non-tilt HA trimer dataset (EMPIAR-10024); and our uniquely collected Na\textsubscript{X} dataset. The relevant details of the data and the corresponding parameters can be seen in Section 1 and Table 1 in the supplementary material. In these experiments, the preprocessing, including motion correction and particle-picking, adhered to the standard cryo-EM SPA workflow. Additionally, the refinement module utilized the widely-used cryo-EM orientation estimation software, CryoSPARC, with its default settings.

Two iterations of cryoPROS were carried out. In the first iteration, the homologous protein's density map was lowpass filtered to $10$\AA, serving as the starting point for the latent volume. The latent volume is updated using the reconstructed density map from raw particles using estimated orientations from the first iteration.  It is shown in Section 2 and Fig. 1 in supplementary material that the impressive fidelity of cryoPROS-generated particles in simulating the signal of raw cryo-electron microscopy (cryo-EM) data. Additionally, we employed post-processing methods, such as EMready~\cite{emready_2023}, as an optional step to address artifacts arising from uneven orientation. We measure the quality of the reconstructed density using the half-map resolution, model-to-map resolution, and Q-scores. These measures collectively serve to validate the density maps generated through cryoPROS quantitatively.

 \subsection{SIM2: eliminating orientation estimation bias.}\label{sec:simulated data}

\begin{figure}[htp]
\begin{center}
\includegraphics[width=\textwidth]{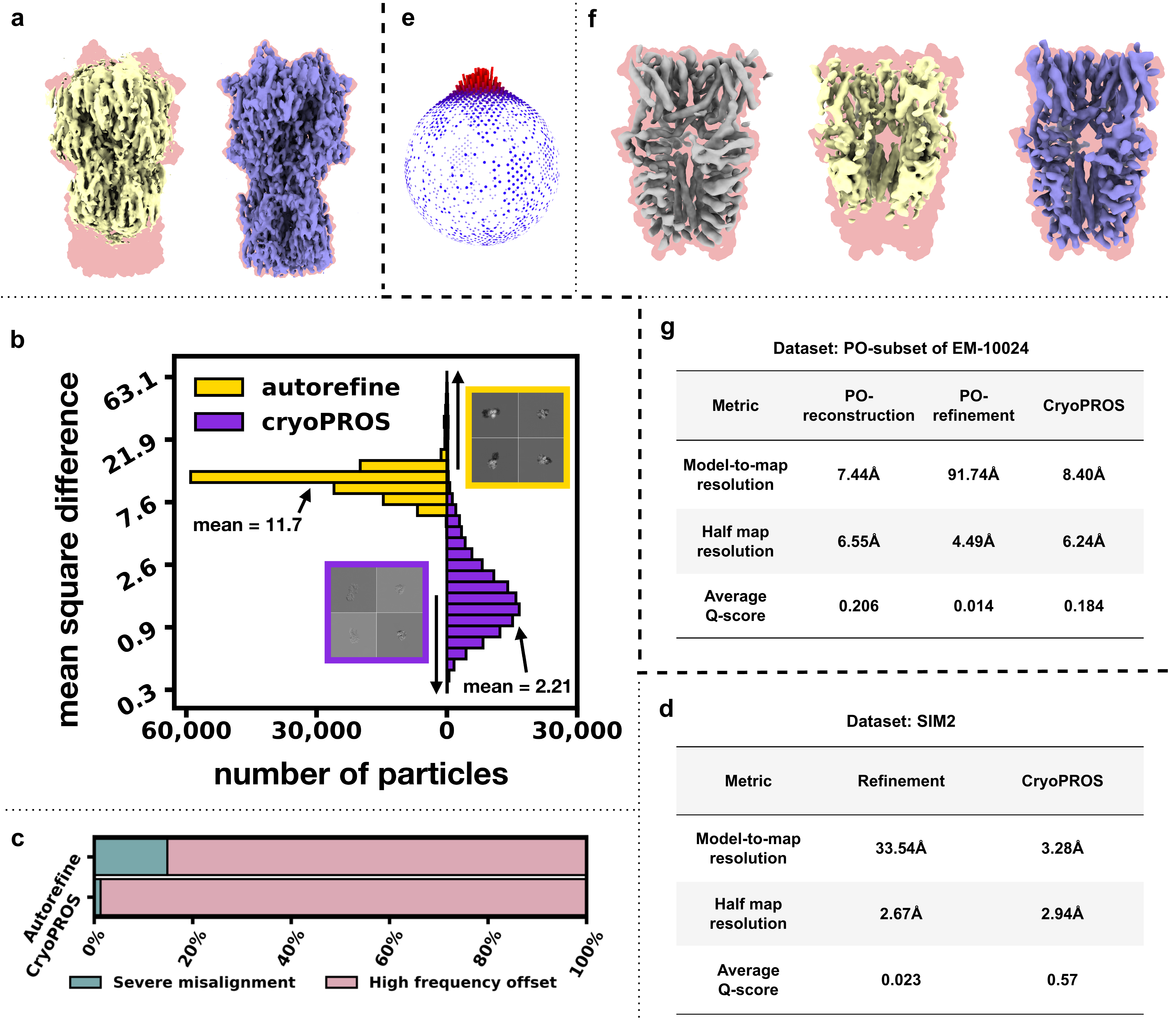}
\caption{\textbf{Avoiding orientation estimation bias with cryoPROS.} \textbf{a}, Reconstructed density maps of SIM2 dataset via autorefinement (in yellow) and cryoPROS (in violet). \textbf{b}, Histograms of MSE the noise-free ground truth projections and the projections derived from the orientation estimated by cryoPROS and CryoSPARC’s autorefine, respectively. Projection differences were highlighted by boxes of yellow and violet, respectively. Means of the MSEs were labeled. \textbf{c}, Proportion of severe misalignment (in green) and high-frequency (in red) offset in autorefinement and cryoPROS misalignment. \textbf{d}, FSC-based resolutions and Q-scores of maps in \textbf{a} are listed. \textbf{e}, Pose distribution of constructed PO-subset of TRPA1. \textbf{f}, Reconstructed density maps of PO-subset of TRPA1 via reconstruction only from fixed pre-estimated and more accurate orientations (in gray), autorefinement (in yellow), cryoPROS (in violet), respectively. Density maps in \textbf{a} and \textbf{f} were superimposed on the envelope of their ground truth atomic models (in red). \textbf{g}, FSC-based resolutions and Q-scores of maps in \textbf{f} are listed.}
\label{fig:recovering bias}
\end{center}
\end{figure}

 We applied cryoPROS to the SIM2 dataset by choosing a homologous protein (PDB ID: 2RFU) with $16\%$ sequence identity to the target protein as the latent volume. The results demonstrated substantial improvements in the density map, particularly in effectively restoring the lost density along the  Z-axis. A model-to-map resolution of 3.28\AA\xspace was achieved by cryoPROS (\cref{fig:recovering bias}a, in violet), as determined by the HA trimer atomic model (PDB ID: 3WHE). This represents a significant improvement compared to the resolution of 33.54\AA\xspace obtained from conventional auto-refinement (\cref{fig:recovering bias}a, in yellow). 

We further compared the pose assignment accuracy between cryoPROS and conventional auto-refinement by computing the mean square difference (MSE) between the noise-free ground truth projection and the projection derived from the orientation estimated by CryoSPARC’s auto-refine (\cref{fig:recovering bias}b). Reduced MSEs of residuals from cryoPROS were observed, indicating overall higher orientation accuracy.  Using residuals greater than $13$ as the threshold for significant orientation errors, cryoPROS decreased the rate of severe misalignments from 14.85\% to 1.27\% (\cref{fig:recovering bias}c). This led to enhanced resolution (\cref{fig:recovering bias}d). These results show the proficiency of cryoPROS in addressing misalignment challenges and achieving superior resolutions.

\subsection{PO-subset of TRPA1: restoring missing density.}\label{sec:TRPA1}

We assessed cryoPROS's ability to restore missing density caused by orientation estimation bias in the case of TRPA1's constructed preferred-oriented dataset. TRPA1 is a $690$kDa prototypical membrane protein and acts as a detector for toxic chemical agents encountered in the environment or produced during tissue damage or drug metabolism \cite{10024}.

We constructed a subset from the deposited raw particle images (EMPIAR-10024~\cite{10024}, containing 43,585 particles) by selecting all top views and a limited number of side views. This resulted in a preferentially oriented subset, referred to as the PO-subset, containing 14,436 particles. The pose distribution of this subset can be found in \cref{fig:recovering bias}e.
Assuming accurate alignment of the raw particle images, we retained the estimated orientation and reconstructed the PO-subset. This process generated a density map, which we named the PO-reconstruction. Despite a slight drop in resolution due to a reduced number of particles, the map remained devoid of artifacts related to preferred orientation (\cref{fig:recovering bias}f, in gray), presenting a holistic structure.

In constrast, the traditional refinement applied to the PO-subset, referred to as the PO-refinement (Fig. \ref{fig:recovering bias}f, shown in yellow), resulted in an unsatisfactory density map. This map suffered from significant density loss along the preferred orientation axis. Furthermore, its quantitative indicators, such as Q-scores and FSC-based resolutions, were notably lower compared to those of the PO-reconstruction (Fig. \ref{fig:recovering bias}g). These observations demonstrate the effects of orientation estimation bias when analyzing structures in preferentially oriented data.

Next, we applied cryoPROS to the PO-subset. Since no deposited structures of homologous proteins were available, we used Alphafold to predict the rat-derived TRPA1 protein (UniProtID: F1LRH9). After the prediction result was lowpass filtered, it served as the first-round input for cryoPROS. The sequence identity between the predicted rat TRPA1 and TRPA1 was $77\%$. After two iterations, cryoPROS generated a more accurate and complete density map of TRPA1 (see \cref{fig:recovering bias}f, right, violet), achieving resolutions of 6.4\AA\xspace  and 8.4\AA\xspace  based on the half-map and map-to-model FSC criteria, respectively (\cref{fig:recovering bias}g). Furthermore, the Q-scores of the density map showed significant improvement (\cref{fig:recovering bias}g). CryoPROS's output resembled the results obtained from PO-reconstruction, indicating its resistance to the density loss caused by the preferred orientation.

\subsection{HA Trimer: achieving near-atomic resolution structure using non-tilted data.}\label{sec:HA trimer}

We utilized cryoPROS to process the non-tilt data with preferential orientation sourced from the EMPIAR repository (EM-10096) and conducted an extensive comparison with the results derived from data using the tilt technique. The tilt data was also obtained from the EMPIAR repository (EM-10097). For the initial round of cryoPROS, we selected a homologous protein (PDB ID: 6IDD) with 47\% sequence identity to the target protein for lowpass filtering as the latent volume. CryoPROS produced uniformly oriented particles after two iterations, with the pose distribution displayed at the bottom of \cref{fig:HA trimer}c. The refinement module then yielded complete and high-resolution results (\cref{fig:HA trimer}a, illustrated in blue). Visually, the quality of the density showed substantial improvement compared to the automatic refinement on non-tilt data, and outperformed the automatic refinement on $40^{\circ}$ tilted data (\cref{fig:HA trimer}b, depicted in pink). However, it is slightly inferior to the state-of-the-art results on tilted data (\cref{fig:HA trimer}b). Achieving such results necessitates complex subsequent refinements at the per-particle level, including multi-round 3D classification, defocus refinement, and Bayesian polishing ~\cite{polish}. In addition, we post-processed the cryoPROS result using EMReady to eliminate the preferred orientation artifact. This post-processing led to a minor improvement in the quality of results (\cref{fig:HA trimer}a, depicted in blue). Throughout the entire process, we refrained from filtering particles based on experience, thereby ensuring a faithful representation of the non-tilt data.

We assessed the quality of density maps using two quantitative metrics: resolution and average Q-score. The resolution measures included the half-map and map-to-model resolutions relative to the HA trimer structure (PDB ID: 3WHE). Q-score evaluations were conducted at the atomic level, main chain level, and side chain level (\cref{fig:HA trimer}f). Notably, the cryoPROS output, following post-processing, demonstrated an impressive model-to-map resolution of 3.90 \AA\xspace and a half-map resolution of 3.27 \AA\xspace. This result validates the effectiveness of cryoPROS in refining preferred orientation data to nearly atomic resolution. Significantly, this achievement was realized without necessitating additional data collection or the complex per-particle operations typically used to mitigate the intrinsic limitations of the tilt strategy.

Furthermore, we selected specific sites in the known atomic model (PDB ID: 3WHE) for density comparison (\cref{fig:HA trimer}d,e). The results demonstrated that the density map generated by cryoPROS was well-preserved, displaying distinct regions with alpha-helical pitch, beta-strand separation, and bulky side chains (\cref{fig:HA trimer}e-i, e-ii, shown in blue). We employed Q-score.bb (backbone, see \cref{fig:HA trimer}e-iii) and Q-score.sc (side chain, see \cref{fig:HA trimer}e-iv) to color the atomic model, which revealed high scores for both the main chain and side chain in the results produced by cryoPROS. In all comparisons, cryoPROS closely mirrored the results derived from the analysis of tilted data. Moreover, in most instances, cryoPROS surpassed the results from tilt-collection-autorefinement (\cref{fig:HA trimer}b, depicted in pink), nearly matching the best results from the current dataset. All these findings suggest that cryoPROS can achieve nearly atomic-resolution density maps from preferentially oriented single-particle datasets.

\begin{figurehere}
\begin{center}
\includegraphics[width=0.9\textwidth]{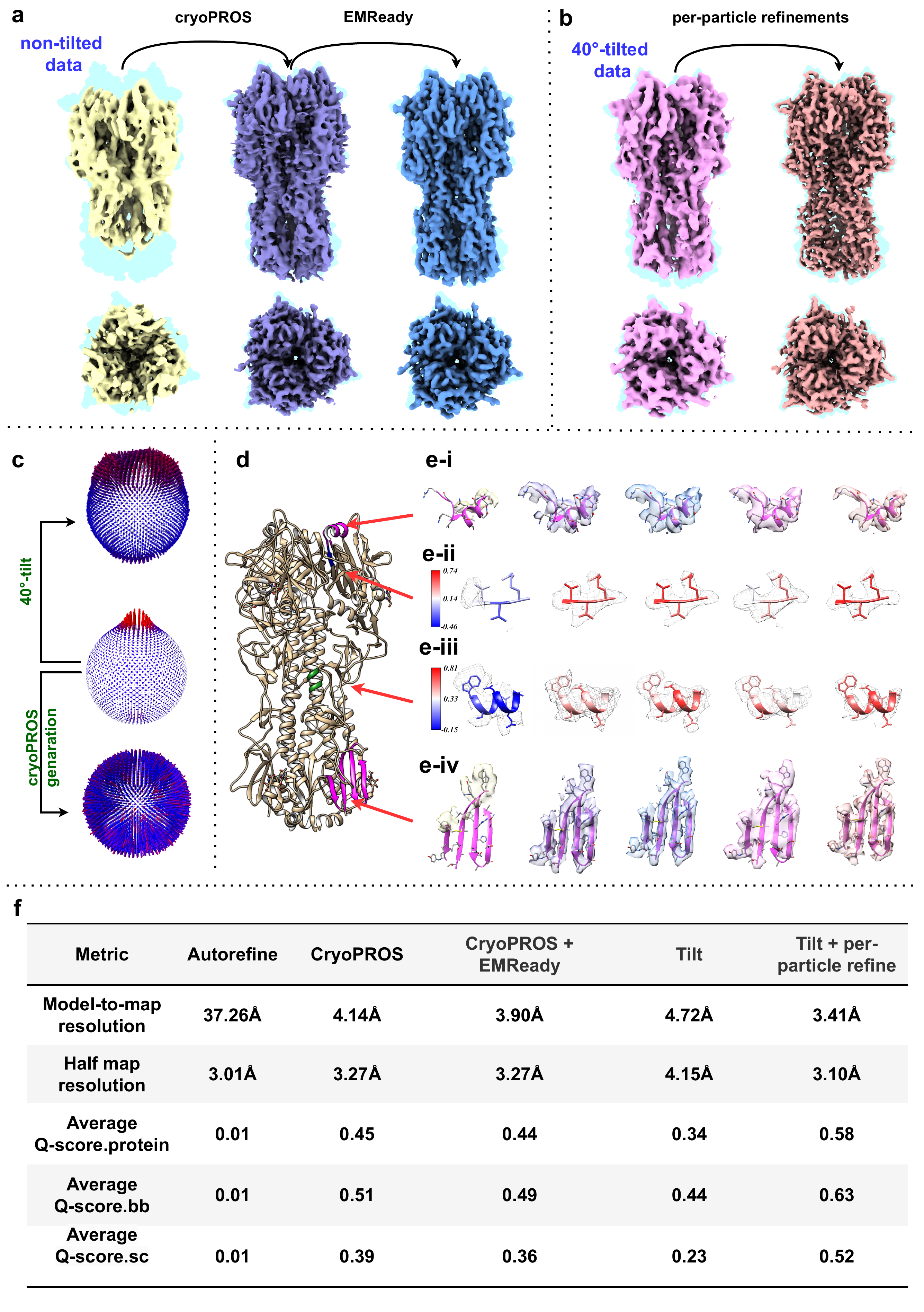}
\caption{\textbf{CryoPROS enables the recovery of near-atomic-resolution information from highly preferentially oriented data of the influenza A HA-trimer.} \textbf{a}, Reconstructed density maps of the non-tilt dataset using: autorefinement (in yellow), cryoPROS (in violet), and cryoPROS with follow-up EMReady postprocessing (in blue). \textbf{b}, Reconstructed density maps of the tilt-collected dataset: Autorefinement (in pink) and state-of-art result (in crimson). Notably, achieving the state-of-the-art result necessitates intricate subsequent refinements at the per-particle level, which encompass multi-round 3D classification, defocus refinement, and Bayesian polishing. Maps in \textbf{a} and \textbf{b} are superimposed on the envelope of the HA-trimer atomic model (PDB ID: 3WHE, in light blue). \textbf{c}, Pose distribution of datasets: $40^{\circ}$-tilted dataset (top), non-tilt dataset (middle), and cryoPROS generated auxiliary particles (bottom). \textbf{d}, Colored parts are selected for close-up in \textbf{e}.  \textbf{e}, Detailed close-ups of selected parts of the density maps shown in \textbf{a} and \textbf{b}, of which order are consistent with. Panel e-i,iv show the regions of alpha-helix and beta-sheet in low transparency. Panels ii and iv show the selected regions in gray mesh style, embedded atomic model is colored by average Q-score.sc value and average Q-score.bb value, respectively.  \textbf{f}, Multiple indicators are used to quantitatively comparel density maps shown in \textbf{a} and \textbf{b}. Q-score evaluations span the atomic level, main chain level, and side chain level,  termed as average Q-score.protein, average Q-score.bb, and average Q-score.sc, respectively.}
\label{fig:HA trimer}
\end{center}
\end{figurehere}

\subsection{Na\textsubscript{X}: Expanding cryoPROS's applicability to membrane protein analysis.}\label{sec:Nax}

Na\textsubscript{X} is an atypical sodium channel involved in regulating sodium homeostasis and other physiological functions, distinguished by its voltage-insensitivity and resistance to tetrodotoxin. It is a potential target for diseases related to sodium balance and other associated conditions~\cite{Nax_2022}. To study this, we gathered a non-tilt dataset containing $411,823$ particle images. Since Na\textsubscript{X} is a membrane protein, it was purified and solubilized using the LMNG detergent for this dataset. This process results in the detergent's visualization as a micelle. Particles in this dataset display a strong preferred orientation, rendering side views or inferior orientations nearly invisible. Furthermore, since they are solubilized by detergent, the sizable micelle presence further complicates orientation estimation. Such disordered densities cannot be overlooked. Thus, we applied cryoPROS-MP, an enhanced version of cryoPROS, to this case.

We used the density map derived from the atomic model of Na\textsubscript{v}1.6 (PDB ID: 8FHD) and applied a lowpass filter to $10$\AA. This map served as the latent volume in the initial iteration of cryoPROS-MP. It shares a $56\%$ sequence identity with the ground truth.

Distinct from cryoPROS approach, cryoPROS-MP's primary enhancement includes a micelle reconstruction step. This captures the latent volume embedded within the micelle, as depicted in pink in \cref{fig:Nax}d. We explore the significance of micelle reconstruction to cryoPROS-MP in greater depth in Supplementary Section 3 and Supplementary Fig.2. When juxtaposed with the conventional autorefinement result (shown in yellow in \cref{fig:Nax}b), the outcome from cryoPROS-MP (displayed in blue in \cref{fig:Nax}b) demonstrates a marked improvement in map quality, particularly emphasizing the clarity of the central transmembrane regions. The atomic model, colored based on the backbone-level average Q-score of the cryoPROS-MP result, highlights the congruence between the map and the model, especially in the central areas (refer to \cref{fig:Nax}d). However, some relatively flexible exterior regions show a less consistent alignment. Using the ground truth atomic model as a benchmark, the cryoPROS result attains a model-to-map resolution of 7.22\AA, with additional validation indicators presented in \cref{fig:Nax}c. Though this resolution is within the medium spectrum, it represents a significant leap over conventional refinement techniques. This resolution constraint largely stems from the total lack of data orientation. In conclusion, cryoPROS-MP serves as an effective expansion of the cryoPROS framework, specifically tailored to membrane proteins. Considering their crucial role in drug discovery, we emphasize the indispensable nature of this enhancement.

\begin{figurehere}
\begin{center}\includegraphics[width=1\textwidth]{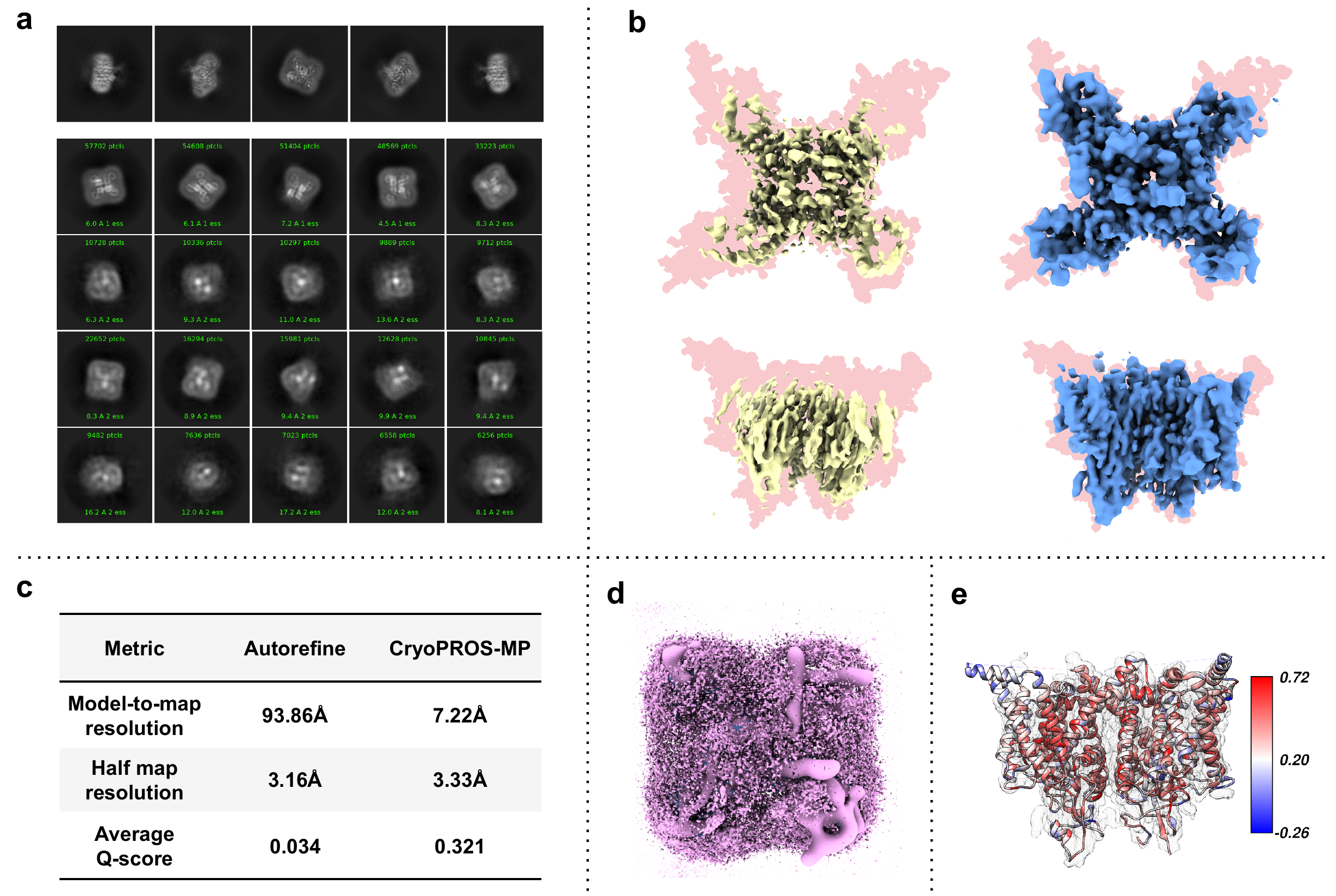}
\caption{\textbf{CryoPROS-MP: Adapting to membrane proteins solubilized in micelles or liquid nanodiscs.} \textbf{a}, The upper row displays projections of the ground truth density map, with orientations chosen at random. The 2D classification averages of the Na\textsubscript{X} dataset, obtained via CryoSPARC, are arranged in a $4 \times 5$ grid below. \textbf{b}, Density maps reconstructed from the non-tilt data using autorefinement (in yellow) and cryoPROS-MP (in blue) are presented. These maps are superimposed on the envelope of the ground truth atomic model (in light red). The side view is displayed in the upper row, while the top view occupies the lower row. \textbf{c},  FSC-based resolutions and Q-scores of maps in \textbf{b} are listed. \textbf{d}, Latent volume for the first iteration of cryoPROS-MP. \textbf{e}, The ground truth atomic model is colored by the backbone-level average Q-score of the resulting cryoPROS-MP map. This map, after being made partially transparent, is superimposed onto the atomic model.}

\label{fig:Nax}
\end{center}
\end{figurehere}
    
\subsection{CryoPROS is unaffected by the potential issue of model bias.}\label{sec:model bias}

Given that cryoPROS requires a lowpass filtered map of a homologous protein as an initial volume for the first iteration, we examined the potential for model bias. Since cryoPROS requires a low-pass filtered map of a homologous protein as an initial volume for its first iteration, it is crucial to evaluate whether model bias influences the resulting map from CryoPROS. CryoRPOS implements three strategies to mitigate this risk.

In its first strategy, cryoPROS undergoes two rounds of iterations. After the first iteration, cryoPROS updates the latent volume, which is reconstructed exclusively from raw particles while excluding the generative ones. Consequently, no orientation estimation directly relies on the projections derived from the homologous protein. Therefore, the risk of model bias is reduced.

In its second strategy, during the training process, the CVAE employs a reconstruction loss that aligns the generated auxiliary particles with raw particles. This step eliminates any detailed features of the homologous protein that might misguide the reconstruction of the target protein. We discovered that the noise generated by the trained CVAE is highly unlikely to overfit into the density maps. In comparison, Gaussian noise is susceptible to overfitting, a phenomenon widely recognized as the ``Einstein from noise'' effect~\cite{einsteinfromnoise_2013}. To further explore this, we conducted the ``Einstein from noise'' experiment. This involved parallel autorefinement of both Gaussian pure noise stacks and CVAE-learned pure noise stacks. We utilized varying degrees of lowpass filtered density maps as references (specifically the ribosome, with PDB ID: 6pcq). The resulting three-dimensional signal density map is essentially a misleading signal derived from noise due to inherent model assumptions. The finer the resolution, the higher the potential for model bias. As depicted in \cref{fig:model bias}b, with a density map lowpass filtered to 7Å serving as a reference, CVAE effectively curbs the interference of model bias. Conversely, Gaussian noise demands more extensive lowpass filtering to achieve a similar bias reduction, as shown in \cref{fig:model bias}b. \cref{fig:model bias}c highlights that the density maps generated by CVAE-learned noise have markedly lower average quality—roughly two orders of magnitude less than maps produced by Gaussian noise. A closer look at the induced density map slices in \cref{fig:model bias}d shows that the CVAE-learned noise is nearly indistinguishable due to its minimal impact from its inferior quality. These tests unequivocally underscore the importance of the cryoPROS generative module in effectively sidestepping model bias.

CryoPROS generates particles by adding noise to the projections of the latent volume. To simulate this complex process, we adopted the CVAE framework. However, a simpler method to generate particles involves adding Gaussian noise directly to the projections by the latent volume. The detail of this method is given in supplementary Section 5.2. Also, we report the results on HA trimer and Na\textsubscript{X} in Supplementary Table 3. We found that applying Gaussian noise does not yield satisfactory results for the Na\textsubscript{X} protein. Although it appears to be effective for the HA trimer, the resultant structure is much closer to the homologous protein than that produced by cryoPROS, suggesting a substantial risk of model bias. This finding validates the fundamental difference between CVAE-learned noise and Gaussian noise. Additionally, we calculated the KL divergence and the noise intensity, as given in \cref{fig:model bias}a. The results demonstrate that the CVAE-learned noise closely mimics the real noise present in raw particles.

In its third strategy, CryoPROS employs a low-pass filtered initial model derived from a homologous protein that shares only low-frequency information with the target protein, intending to retain only the characteristic signals common to both the homologous protein and the target protein. This methodology is broadly endorsed within the cryo-EM field as a means to circumvent the issue of model bia~\cite{sorzano_bias_2022}. For further validation of its effectiveness in the context of CryoPROS, we conducted an ablation study. This study assesses CryoPROS's robustness against variations in the threshold of the low-pass filter and the sequence similarity between the target protein and the homolog.

We selected a homologous protein with a lower sequence identity (PDB ID: 2RFU, with only $16$\% sequence identity), denoted as cryoPROS-2RFU. The results of this experiment, illustrated in blue in \cref{fig:model bias}e, largely align with the previous experiment based on the homologous protein 6IDD (referred to as cryoPROS-6IDD, results shown in blue-green in \cref{fig:model bias}e). This consistency suggests that the effectiveness of cryoPROS is not heavily reliant on the choice of the homologous protein. The atomic models of the homologous proteins used in these two experiments, as well as the latent volume from the first round of cryoPROS, are displayed in \cref{fig:model bias}e. In addition, our examination of real data (Na\textsubscript{X}), detailed in Supplementary Section 4.1 and Supplementary Table 2, demonstrates the consistent stability of outcomes when using different homologous proteins. Furthermore, our exploration in Supplementary Section 4.2 and Supplementary Fig.3 delves into the impact of distinct lowpass parameters of the initial latent volume on cryoPROS outcomes. Our findings suggest that cryoPROS maintains a low risk of introducing model bias when various lowpass filter parameters are utilized. However, regarding the accuracy, the current configuration of cryoPROS with a 10\AA\xspace lowpass is the default choice. 

We subsequently compared the results of cryoPROS-6IDD and cryoPROS-2RFU with the currently best results (achieved via Tilting+per-particle refinements) for structure refinement based on the atomic model 3WHE. We evaluated the correspondence between the three refined structures using TMscore and RMSD. Both sets of cryoPROS results were found to be remarkably close to the current best results and significantly divergent from the homologous proteins. Similar outcomes were observed in other datasets (refer to Supplementary Section 5.3 and Supplementary Fig.4 for more details). As shown in \cref{fig:model bias}f, the positional relationship of atomic models from the three groups at different sites confirms that cryoPROS's results align more closely with the best results procured by tilting, rather than being nearer to the homologous proteins.

Homologous proteins and target proteins are expected to share similar low-frequency features while differing in high-frequency local features. We computed the FSC curves of the atomic model-to-map for homologous proteins, target protein ground truth map, and cryoPROS result, and we determined the resolution based on this FSC curve (see \cref{fig:model bias}g). Taking homologous protein 6IDD as an example, we found that the correlation between the density map obtained by cryoPROS and the homologous protein is lower than that of the target protein in each frequency band. 
By setting a threshold correlation of 0.5 (corresponding to 8.59\AA\xspace), we delineated the common characteristic frequency band of the two proteins (lower than 8.59\AA\xspace) and the unique characteristic frequency band of the homologous protein (higher than 8.59\AA\xspace). The correlation of the cryoPROS result fell below 0.5 at 20.96\AA\xspace, which is in the low-frequency band. This suggests that cryoPROS successfully avoids incorporating specific local details from homologous proteins.

Finally, we conducted a series of local refinement experiments subsequent to cryoPROS (see \cref{fig:model bias}h), which validated the stability of cryoPROS's results.

\begin{figurehere}
\begin{center}
    
\includegraphics[width=1\textwidth]{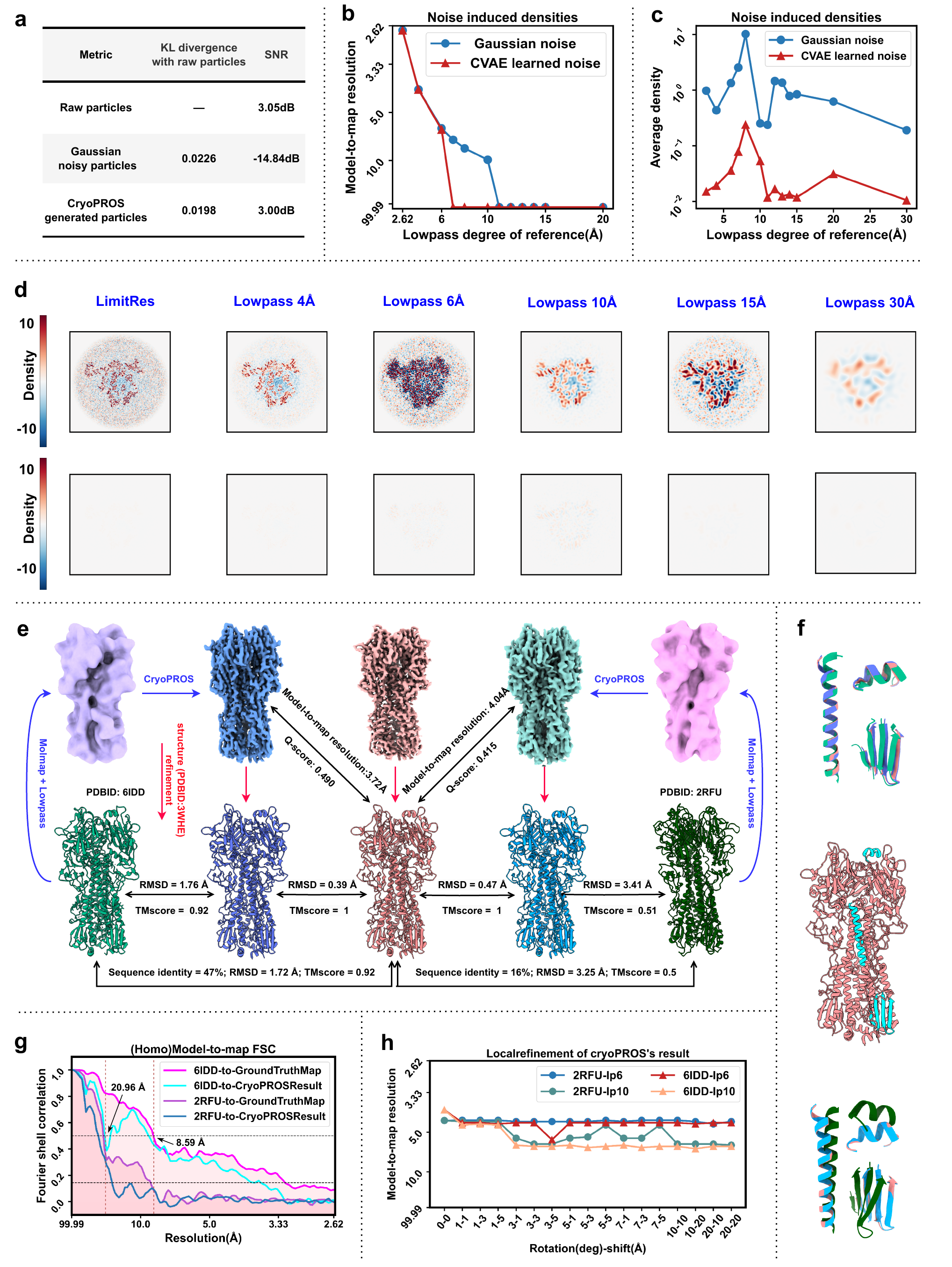}
\caption{\textbf{CryoPROS is unaffected by the potential issue of model bias.} \textbf{a}, Evaluation of the similarity between and generated noisy particles (including CVAE noisy and Gaussian noisy), employing KL divergence and SNR as the evaluation metrics, see Supplementary Section 5.1 for calculation details. \textbf{b}, Model-to-map FSC resolution between the alignment reference and the ``phantom'' density maps obtained from ``Einstein from noise'' three-dimensional experiment: perform autorefinement on Gaussian pure noise stacks and CVAE pure noise stacks with different degrees of lowpass filtered density maps as references (ribosome, PDB ID: 6pcq). \textbf{c}, Comparison of the average density of maps obtained in the ``Einstein from nois'' three-dimensional experiment. \textbf{d}, Slice comparison of the maps obtained the“Einstein from noise” three-dimensional experiment. \textbf{e}, CryoPROS experiments using homologous proteins. Two sets of experiments were conducted using homologous proteins different in sequence identities: cryoPROS-6IDD and cryoPROS-2RFU. Density map results (top row) and associated atomic model (bottom row) are presented, along with structure refinements based on the atomic model 3WHE. The homologous proteins utilized in the two experimental groups are introduced on the left and right sides: 6IDD (atomic model shown in light green, lowpass filtered density in light purple); 2RFU (atomic model shown in dark green, lowpass filtered density in pink). The middle three densities are as follows: cryoPROS-6IDD results (displayed in dark blue), best results from the Tilt strategy (shown in red), and cryoPROS-2RFU experimental outcomes (in blue-green), their structure refinement results are shown below, respectively. RMSD and TM-score were utilized to assess similarity and demonstrate cryoPROS's low model bias risk property.  \textbf{f}, Alignment and comparison of atomic models. All atomic models in \textbf{g} were aligned and compared, focusing on the selected region (middle). The upper and lower sections display relevant results from the two groups of cryoPROS-6IDD and cryoPROS-2RFU experiments, respectively. \textbf{g}, (Homologous Protein) model-to-map FSC curve. \textbf{h}, Follow-up local refinement experiment on cryoPROS, using different local search parameters, with few changes observed. The two values in the figure's legend correspondingly signify the homologous protein employed in the cryoPROS experiment and the low-pass parameters utilized in the local refinement experiment. To illustrate, 6IID-lp6 signifies the application of a 6-\AA\xspace low-pass filter on the 6IID-cryoPROS resulting map as the reference of local refinement and then local search the orientation on the basis of that is 6IID-cryoPROS estimated. The x-axis illustrates the extent of rotation and displacement searches.
}
\label{fig:model bias}
\end{center}
\end{figurehere}

\section{Conclusion and Discussion}

We presented cryoPROS, a novel solution that overcomes the key pain point hindering high-resolution analysis of preferentially-oriented data, which is the gradually accumulated orientation estimation bias. CryoPROS corrects this bias through a pure calculation method without the need for additional sample preparation, data acquisition methods, or computational refinement steps, thus reducing the experimental burden. To verify its effectiveness and reliability, we conducted experiments using several datasets and successfully applied cryoPROS to non-tilt data single particle analysis of Na\textsubscript{X} and HA trimer, achieving high-resolution results comparable to the Tilt strategy. Furthermore, extensive experiments demonstrate cryoPROS's ability to minimize model bias risk effectively. These findings highlight the significant potential of cryoPROS as a valuable tool for advancing single-particle cryo-EM structural analysis of frozen samples. Moreover, our method encourages open discussions and welcomes further exploration of the following aspects.

\textbf{The source of the first round of latent volume.} In the standard practice of cryoPROS, the first round of latent volume input is derived from the lowpass filtered homologous protein of the target protein. The homologous protein provides a reliable low-frequency prior to the initial CVAE learning, and its low frequencies are consistent with those of the target protein, effectively improving the quality of CVAE generation. While we have observed cryoPROS's stable performance with different sequence identities of homologous proteins, the importance of the homologous protein cannot be overlooked. In cases where homologous proteins are lacking, we plan to explore alternative methods, such as predicting an initial model using powerful structure prediction methods like Alphafold2\cite{Alphafold}, as demonstrated in Section\ref{sec:TRPA1}. Another approach is to post-process traditional reconstruction results to obtain a lowpass filtered but relatively complete structure, serving as the first-round input for cryoPROS.

\textbf{Potential applications of conditional generative model.} In cryoPROS, we have designed a CVAE model capable of generating realistic particles corresponding to any specified pose. The noise distribution in the synthesized particles has been verified to be strikingly similar to that of real noise, presenting a lower model bias risk compared to Gaussian noise. In the field of cryo-EM, both single-particle analysis and electron tomography face the fundamental challenge of managing ultra-low SNR in observed data, thereby emphasizing the critical importance of precise noise modeling. One potential application of this generative model is to utilize the synthesized particles to mitigate the difficulties associated with observed data having imbalanced orientations.
For example, in 3D classification, generating particles corresponding to less favored categories could potentially improve the process of particle classification. Moreover, the conditional generative model could be employed to synthesize paired training data for downstream tasks~\cite{9924527}. In signal recovery, a dataset composed of synthesized pairs of clean and noisy particles can be harnessed for training a denoising network, thereby enhancing the signal in the raw particles. Similarly, in pose estimation, the synthesis of particle-pose pairs facilitates the training of a pose estimation network dedicated to determining the orientations of the original particles

\textbf{Mixture of preferred orientation and other issues.} In real-world scenarios, the preferred orientation problem is often intertwined with other challenges, amplifying the complexity of finding a solution. For instance, membrane proteins embedded in detergent micelles or lipid nanodiscs present additional difficulties, and we have extended the cryoPROS setting to address such cases in Section\ref{sec:Nax}. More generally, the complexities of cryo-EM analysis lead to a combination of various issues, such as heterogeneity, dynamics, and orientation difficulty in the case of the small protein. Successfully employing cryoPROS to solve these hybrid problems may require further refinement and practical experimentation.

\textbf{Post-processing method.} In this study, we have divided the preferred orientation problem into two components: orientation estimation bias and anisotropic reconstruction. The core focus of cryoPROS is to address the orientation estimation bias, which is considered more critical. However, in high-resolution structure determination, artifacts resulting from anisotropic reconstruction also will impact structure observation. To address this issue, we introduced EMReady for post-processing the output of cryoPROS, especially in the case of the HA trimer (see Section\ref{sec:HA trimer}). Nevertheless, existing post-processing methods like EMReady may not specifically target artifacts related to the dominant orientation density. As a future direction, we aim to develop dedicated tools that can better handle and resolve these artifacts to further enhance the effectiveness of cryoPROS in high-resolution structure analysis.

\section*{Acknowledgements}
	
This work was supported by the National Key R\&D Program of China (No.2021YFA1001300) (to C.B.), the National Natural Science Foundation of China (No.12271291) (to C.B.), the Beijing Frontier Research Center for Biological Structure (to M.H.), Shenzhen Medical Academy of Research and Translation (to M.H.) and the National Natural Science Foundation of China (No.12071244) (to Z.S.). We are grateful to Dr. Gaoxingyu Huang for his valuable discussions and to Prof. Yigong Shi for his support in data analysis.

\section*{Data availability}
The data that support this study are available from the corresponding authors upon request. Three raw datasets analyzed in this study were downloaded from the EMPIAR repository (EMPIAR-10096, EMPIAR-10096, EMPIAR-10097). SIM1 and SIM2 were generated by the {\texttt relion\_project} module within Relion. Dataset Na\textsubscript{X} was collected in-house and will be deposited into EMPIAR in the near future. Structures for the initial latent volume of cryoPROS were downloaded from the Protein Data Bank (PDB ID: 2RFU, 6IDD, 5XL8, 7XM9, 8FHD, 6AGF) or downloaded from AlphaFold Protein Structure Database (UniProt ID: F1LRH9). The structure for validation was based on the HA trimer atomic model (PDB ID: 3WHE) and the TRPA1 ion channel atomic model (PDB ID: 3J9P).

\section*{Code availability} 
CryoPROS will be open-source upon publication and is also available upon request during the review process.

\section*{Contribution} 
H.Z, D.Z, C.B., M.H., and Z.S. initiated the project. H.Z., D.Z., C.B. and M.H. developed CryoPROS and carried out testing. H.Z. and D.Z. analyzed the data. Q.W. and N.Y. collected the Na\textsubscript{X} dataset. H.Z., D.Z., C.B., M.H., and Z.S. wrote the manuscript.

\section*{Competing interests}
All other authors declare no competing interests.

\bibliography{1_manuscript.bib}
\section*{Methods}
\subsection*{The Generative module}\label{sec:model_formulation}
The generative module using the CVAE model aims to generate realistic particles, which is the key component in cryoPROS. Let $\{\vx_i\}_{i=1}^N$ be the set of raw particles, we can obtain the estimated orientation parameters $\{\phi_i\}_{i=1}^N$ and CTF parameters $\{\psi\}_{i=1}^N$ from cryoSPARC and adopt the imaging model as
\begin{equation}
	\vx_i = \vC(\psi_i) P(\phi_i) \vV_{latent} + \vn_i,\quad i=1,2,\cdots,N,
\end{equation}
where $\vC$ is the CTF operator depends on $\psi_i$,  $P(\phi)$ represents the projection operator for the pose $\phi$, and $\vn_i$ represents the noise. It is noted that the noise term $\vn_i$ contains both random error and the model error induced by the estimation error of parameters. Define $\vTheta=\{\psi,\phi,\vV_{latent}\}$ to be the set of imaging parameters, the CVAE model learns a generative model that maximizes the conditional likelihood $p(\vx\lvert\vTheta)$, which is equivalent to the unpaired data modeling problem~\cite{9924527} between the latent projection $\vv = \vC(\psi) P(\phi) \vV_{latent}$ and raw particles. 

Since $p(\vx\lvert\vTheta)$ is difficult to optimize directly, the CVAE model involves maximizing a lower bound of $\log p(\vx\lvert \vTheta)$, known as the conditional Evidence Lower BOund (cELBO). The negative cELBO is given by
\begin{equation*}\label{loss}
 \text{Loss}:=\left(\EE_{q(\vz\lvert\vx, \vTheta)} \log p(\vx \lvert \vz, \vTheta) - \KL(q(\vz\lvert\vx,\vTheta) \| p(\vz\lvert\vTheta))\right),
\end{equation*}
forming the loss function to train our model. Here, $\vz$ is the latent variable, $q(\vz\lvert\vx, \vTheta)$ is the inference model, $p(\vz\lvert\vTheta)$ is the prior model, and $p(\vx \lvert \vz,\vTheta)$ is the generative model. All these models are parametrized by deep neural networks. After the training stage, inputting an imaging parameter $\vTheta$, the CVAE model samples the latent variable $\vz$ through the prior model $p(\vz\lvert\vTheta)$, and subsequently, $\vz$ is transformed into a synthetic particle through the generative model $p(\vx \lvert \vz,\vTheta)$. Furthermore, to enhance the representation capability of the traditional VAE model, a hierarchical structure is employed to address the issue of generating unrealistic and blurry samples in the single-layer model. In particular, we assume the latent variable has $L$ stochastic layers: $\vz = (\vz^1,\dots,\vz^L)$, and a top-down structure~\cite{sonderby2016ladder} is adopted for the inference and generation process in the CVAE model. Please see Section 6 in the supplementary material for the detailed network architecture. 

The CVAE model undergoes training for 20,000 iterations using the Adam optimizer~\cite{kingma2014adam}. We set the learning rate to $1 \times 10^{-4}$ and the batch size to 8. The model architecture includes 15 hierarchical layers. To mitigate the issue of posterior collapse, we employ the KL annealing method~\cite{bowman2016generating}, implementing a linear annealing scheme during the first 10,000 iterations.

\subsection*{The refinement module}
After the generative module, the generated particles are fed into a second module: the refinement module, which aims to optimize the 3D density map reconstruction of preferred orientation data. Generated particles are used for balance pose distribution, helpful for each round of the alignment of the particles, that are of core for structure determination of preferentially oriented single particle samples but problematic to contribute positively in conventional refinement protocol. Specifically, a two-step operation on any popular user-chosen single-particle cryo-EM software, without additional programs: perform autorefinement on a combination of raw and generated particle stacks; then reconstruct raw stack only within fixed refined orientations, obtaining refined pose parameters and 3D density map finally, all of them will be input generative module for improving the generated particles quality and minimize the model bias introduced by homologous proteins to some extent.

It is worth mentioning that all datasets analyzed in this paper consist of preprocessed particle stacks. The initial orientation parameters needed for the first round of cryoPROS can be derived by conducting standard initialization reconstruction and autorefinement on the particle stack, in which intricate preprocessing of raw cryo-EM data is not required. To ensure a thorough understanding, we offer a concise overview of the conventional single-particle preprocessing method. This will enable readers to begin with raw data processing and seamlessly apply cryoPROS in their research. The preprocessing steps consist of several key components. First, apply MotionCor2 to correct the beam-induced motion in the micrograph movie stacks. Next,  determine the contrast transfer function parameters for each micrograph by GCTF. Then employ CryoPARC for automatic particle picking in micrographs, and extract the particle stack based on the picked particle coordinates. Subsequently, perform a 2D classification in cryoSPARC, and select well-performing 2D-averaged particles for initial model generation and refinement. Only one optional software/method is introduced here, as others yield similar effects and won't be detailed further.

%

%

\end{document}


\title[Article Title]{Supplementary for Addressing preferred orientation in single-particle cryo-EM through AI-generated auxiliary particles}


\author[1]{\fnm{Hui} \sur{Zhang}}
\equalcont{These authors contributed equally to this work.}

\author[2]{\fnm{Dihan} \sur{Zheng}}
\equalcont{These authors contributed equally to this work.}

\author[6,7,8]{\fnm{Qiurong} \sur{Wu}}
\author[5,6,7,8,9]{\fnm{Nieng} \sur{Yan}}

\author*[2,3]{\fnm{Zuoqiang} \sur{Shi}}\email{zqshi@tsinghua.edu.cn}
\author*[5,6]{\fnm{Mingxu} \sur{Hu}}\email{humingxu@smart.org.cn}
\author*[2,3,4]{\fnm{Chenglong} \sur{Bao}}\email{clbao@mail.tsinghua.edu.cn}
\affil[1]{\orgdiv{Qiuzhen College}, \orgname{Tsinghua University}, \orgaddress{\city{Beijing}, \country{China}}}
\affil[2]{\orgdiv{Yau Mathematical Sciences Center}, \orgname{Tsinghua University}, \orgaddress{\city{Beijing}, \country{China}}}
\affil[3]{\orgname{Yanqi Lake Beijing Institute of Mathematical Sciences and Applications}, \orgaddress{\city{Beijing}, \country{China}}}
\affil[4]{\orgname{State Key Laboratory of Membrane Biology}, \orgdiv{School of Life Sciences}, \orgdiv{Tsinghua University}, \orgaddress{\city{Beijing}, \country{China}}}
\affil[5]{\orgname{Shenzhen Medical Academy of Research and Translation (SMART)}, \orgaddress{\city{Shenzhen}, \country{China}}}
\affil[6]{\orgname{Beijing Frontier Research Center for Biological Structure (Tsinghua University)}, \orgaddress{\city{Beijing}, \country{China}}}
\affil[7]{\orgdiv{Tsinghua-Peking Joint Center for Life Sciences}, \orgdiv{Tsinghua University}, \orgaddress{\city{Beijing}, \country{China}}}
\affil[8]{\orgdiv{School of Life Sciences}, \orgdiv{Tsinghua University}, \orgaddress{\city{Beijing}, \country{China}}}
\affil[9]{\orgdiv{Department of Molecular Biology}, \orgdiv{Princeton University}, \orgaddress{\city{Princeton, NJ}, \country{USA}}}


\maketitle

\section{Dataset information}

In this study, we evaluate the performance and attributes of the newly developed cryoPROS method, utilizing a diverse set of datasets that include both simulated and real-world cryo-electron microscopy (cryo-EM) data. Notably, the SIM2, PO-subset, HA trimer, and $\text{Na}_{\text{X}}$ datasets present challenges associated with preferred orientation issues. A detailed evaluation of cryoPROS on these specific datasets is thoroughly discussed in Sections 2.3-7, shedding light on its effectiveness in various scenarios. Comprehensive details of all datasets are carefully cataloged in Table~\ref{table:data}, which includes pertinent information such as data source, particle count, target protein, and the resolution achieved by cryoPROS, among other factors.

\begin{sidewaystable}[h]
\renewcommand{\arraystretch}{1.3}
\centering
\begin{tabular}{{ccccccc}}
\hline
\rowcolor[gray]{0.95}
Datasets &
  SIM1 &
  SIM2 &
  EM-10024 &
  \begin{tabular}[c]{@{}c@{}}PO-subset  of  \\ EM-10024\end{tabular} &
  HA Trimer &
  Na\textsubscript{X} \\ 
Data Source &
  Simulation &
  Simulation &
  \begin{tabular}[c]{@{}c@{}}EMPIAR  \\ repository\end{tabular} &
  \begin{tabular}[c]{@{}c@{}}Manual  \\ selection\end{tabular} &
  \begin{tabular}[c]{@{}c@{}}EMPIAR  \\ repository\end{tabular} & Collected in-house
   \\ 
\rowcolor[gray]{0.95} 
Particles &
  130,000 &
  130,000 &
  43,585 &
  14,436 &
  130,000 &
  411,520 \\ 
\begin{tabular}[c]{@{}c@{}}Preferred \\ orientation\end{tabular} &
  No &
  Yes &
  Yes &
  Yes &
  Yes &
  Yes \\ 
\rowcolor[gray]{0.95}
Target Protein &
  \begin{tabular}[c]{@{}c@{}}HA in H3N2  \\ influenza \\ A viruses\end{tabular} &
  \begin{tabular}[c]{@{}c@{}}HA in H3N2  \\ influenza \\ A viruses\end{tabular} &
  \begin{tabular}[c]{@{}c@{}}TRPA1 ion \\  channel\end{tabular} &
  \begin{tabular}[c]{@{}c@{}}TRPA1 ion  \\ channel\end{tabular} &
  \begin{tabular}[c]{@{}c@{}}HA in H3N2  \\ influenza \\ A viruses\end{tabular} &
  Na\textsubscript{X} \\ 
\begin{tabular}[c]{@{}c@{}}PDB ID  \\ (target protein)\end{tabular} &
  3WHE &
  3WHE &
  3J9P &
  3J9P &
  3WHE &
  None \\ 
\rowcolor[gray]{0.95}
\begin{tabular}[c]{@{}c@{}}Latent volume source  \\ for cryoPROS iter1\end{tabular} &
  None &
  \begin{tabular}[c]{@{}c@{}}Homolog \\ Protein\end{tabular} &
  None &
  \begin{tabular}[c]{@{}c@{}}Alphafold2 \\  predicted\end{tabular} &
  \begin{tabular}[c]{@{}c@{}}Homolog \\ Protein\end{tabular} &
  \begin{tabular}[c]{@{}c@{}}Homolog\\  Protein\end{tabular} \\ 
\begin{tabular}[c]{@{}c@{}}Latent volume  \\ for  cryoPROS iter1\end{tabular} &
  None &
  \begin{tabular}[c]{@{}c@{}}HA in\\ Influenza \\ B virus\end{tabular} &
  None &
  Rat TRPA1 &
  \begin{tabular}[c]{@{}c@{}}HA in H7N9  \\ influenza \\ A viruses\end{tabular} &
  Nav1.6 \\ 
  \rowcolor[gray]{0.95}
\begin{tabular}[c]{@{}c@{}}Resolution obtein-\\ -ed by cryoPROS\end{tabular} &
  None &
  3.28\AA &
  None &
  8.4\AA &
  3.90\AA &
  7.22\AA \\ \hline

\end{tabular}
\caption{\textbf{Attributes of simulated and authentic cryo-EM datasets appeared in the main text}}
\label{table:data}
\end{sidewaystable}

\section{Particles generated by cryoPROS}

\begin{figure}[h]
\centering 
\includegraphics[width=0.9\textwidth]{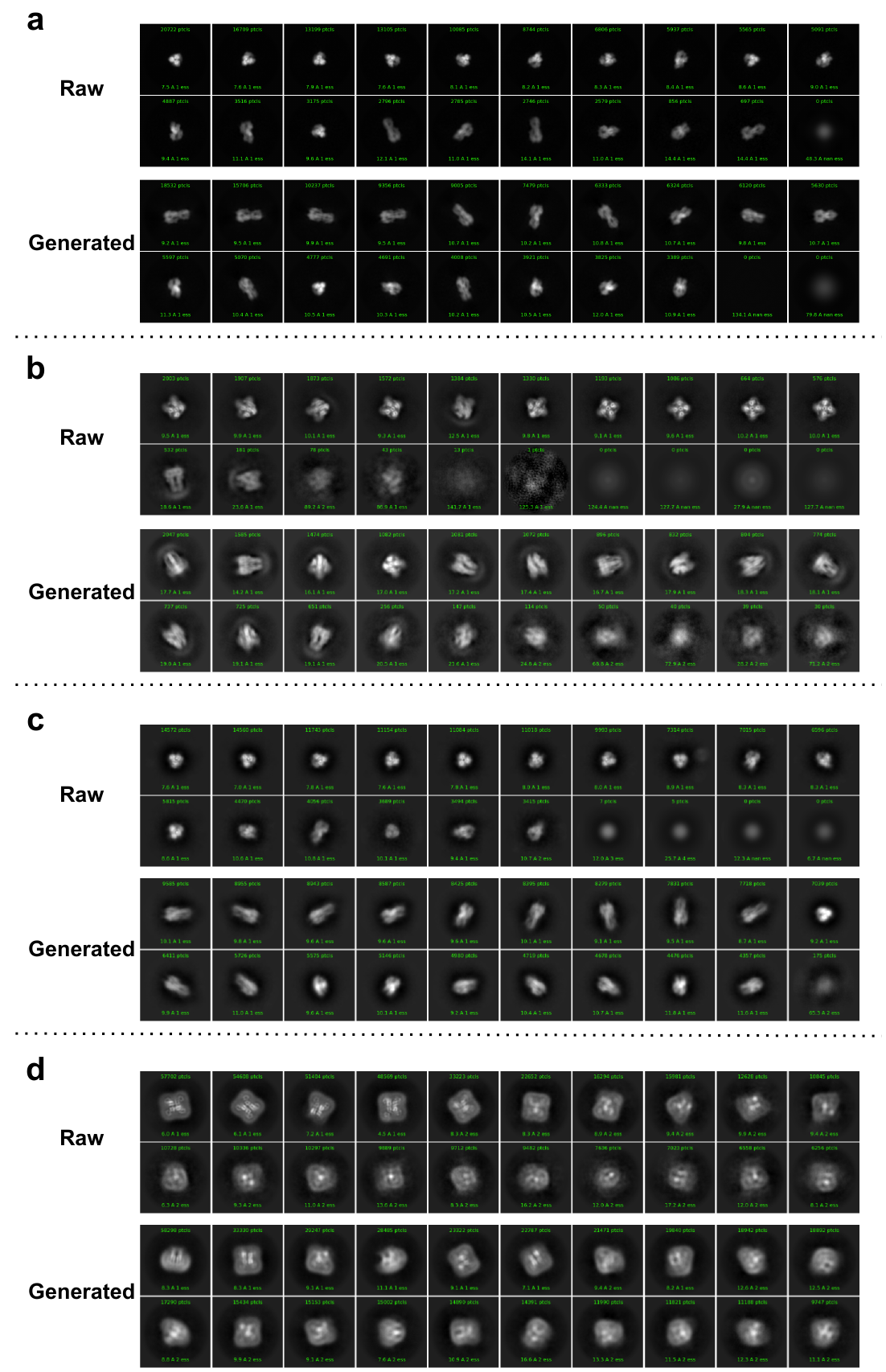}
\caption{\textbf{Comparative analysis of raw and generated particles.} Averaged 2D classification comparison illustrating the similar signal and different pose distributions between raw particle data (upper row) and generated particles (lower row) across preferred oriented datasets, panels a to b, including SIM2, PO-subset of EM-10024, HA trimer, and $\text{Na}_{\text{X}}$.}
\label{sup_fig:generated data}
\end{figure}

The efficacy of CryoPROS relies on leveraging the powerful representational capabilities of deep learning, specifically through the use of the Conditional Variational Autoencode (CVAE) model for training and generating particle data. This strategic approach aims to rectify the distribution of particle poses within datasets that have a preferred orientation, thereby facilitating the execution of the cryoPROS refinement module. The quality of the generated particles inherently impacts subsequent operational efficiency and the feasibility of achieving credible results. \cref{sup_fig:generated data} illustrates the impressive fidelity of cryoPROS-generated particles in mimicking the signal characteristics of authentic cryo-electron microscopy (cryo-EM) data. It's noteworthy that these generated particles exhibit a uniform distribution of orientations.

\section{Importance of micelle reconstruction in cryoPROS-MP}

\begin{figure}[h]
\centering 
\includegraphics[width=0.9\textwidth]{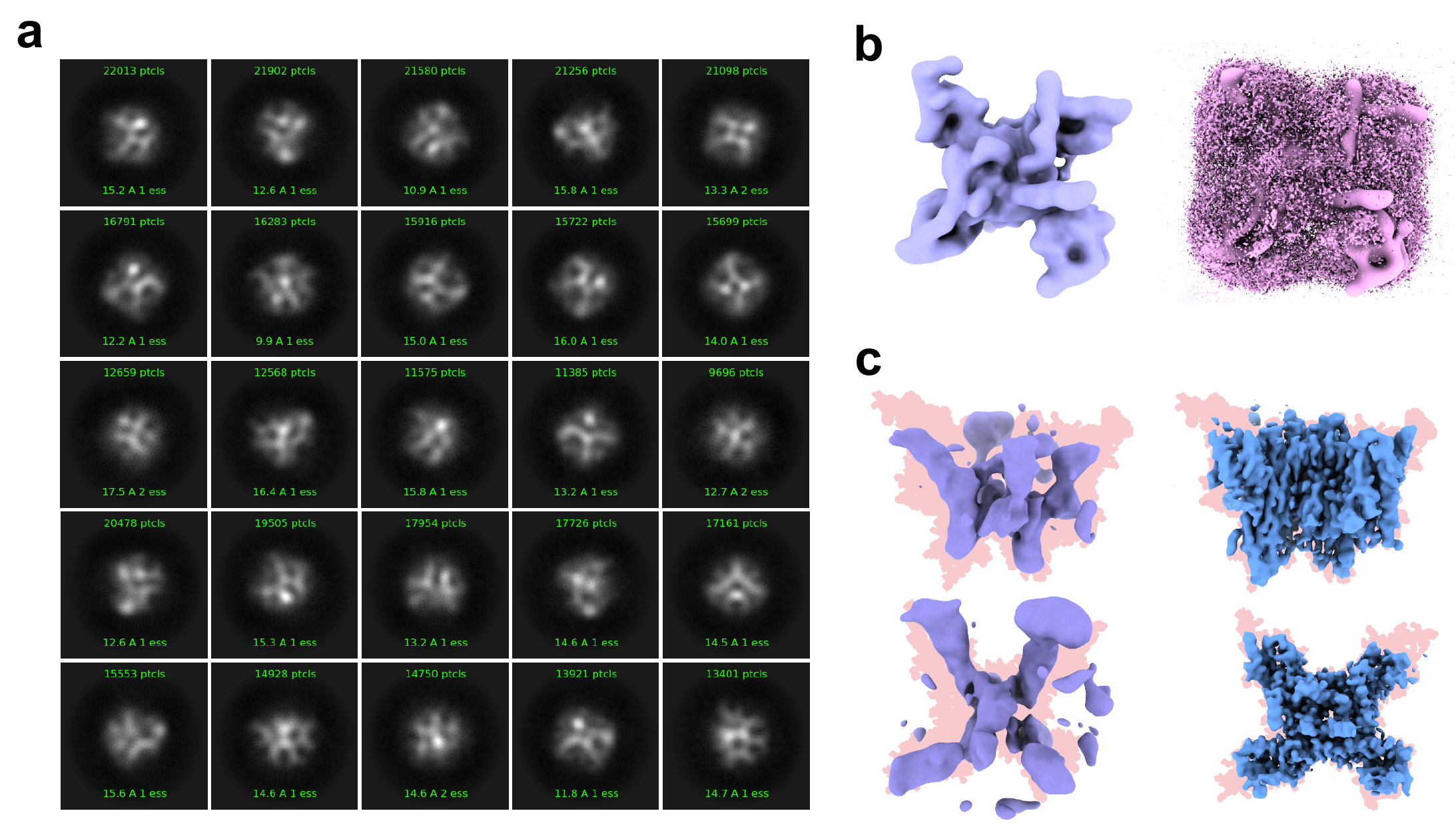}
\caption{\textbf{Necessity of Micelle Reconstruction Step in CryoPROS-MP.} \textbf{a}, Two-dimensional classification average obtained by excluding the micelle reconstruction step in CryoPROS-MP. The comparison illustrates the effects of this omission on particle generative quality. \textbf{b}, Visualization of latent volumes before and after the micelle reconstruction step. \textbf{c}, Resulting map comparison between CryoPROS (left) and CryoPROS-MP (right), highlighting the improvements achieved by the ``micelle reconstruction" module in enhancing structural accuracy and detail.}
\label{sup_fig:nax}
\end{figure}

We present the results derived from the application of cryoPROS to the membrane protein $\text{Na}_{\text{X}}$. For this study, the samples were suitably embedded in a micelle. The two-dimensional classification averages of particle images obtained through cryoPROS are given in \cref{sup_fig:nax}a. Additionally, we provide the reconstructed density maps produced by both cryoPROS and cryoPROS-MP methods, displayed in~\cref{sup_fig:nax}b and \cref{sup_fig:nax}c, respectively. A comparative analysis of these results underscores the enhanced performance offered by cryoPROS-MP. This superior performance firmly substantiates the incorporation of the additional "micelle reconstruction" module within this method.
These promising findings not only validate the efficacy of cryoPROS-MP but also open a new avenue for its application in tackling more complex cases in the future.

\section{Robustness analysis of cryoPROS and cryoPROS-MP}

Besides the input raw particle images, both cryoPROS and cryoPROS-MP depend on an initial latent volume. Currently, this volume is derived by lowpass-filtering the density map of a homolog. Consequently, it's pivotal to examine the robustness of cryoPROS and cryoPROS-MP in relation to this initial latent volume. Specifically, we must assess whether performance is influenced by the choice of homologous protein and the lowpass filter's cutoff frequency. In the subsequent sections, we analyze the impact of different homologous proteins and various lowpass filter cutoff frequencies on the results.

\subsection{Effect of the choice of homologous protein.}

Our experimental design involved the deployment of three distinct homologous proteins within both the HA trimer and Na\textsubscript{X} datasets. For the HA trimer, the abundance of readily available homologous proteins provided an intriguing opportunity for investigation. We selected homologous proteins with varying degrees of sequence identity—67\%, 47\%, and 17\% relative to the target model, in order to thoroughly probe the impact of sequence divergence on cryoPROS performance. Remarkably, despite these differing sequence identities, the quantitative results of the reconstructed density maps in Table~\ref{table:diff_homo} exhibit comparable performance.

In contrast, for the Na\textsubscript{X} dataset, the lack of homologous proteins with various sequence similarities posed a major challenge. As a result, our methodology involved choosing homologous proteins with relatively closer sequence identities to the target model. Despite the limited range of homologous proteins, our results maintained similar model-to-map resolutions for the reconstructed density maps~(see Table~\ref{table:diff_homo}). The above two experiments establish the robustness of the proposed cryoPROS and the enhanced version cryoPROS-MP with respect to homologous proteins.

\begin{table}
	\setlength{\tabcolsep}{8pt}
	\renewcommand{\arraystretch}{1.5}
	\centering	
	\begin{tabular}{cccccc}
		\hline
		\rowcolor[gray]{0.95}

        &
		PDB ID & Half-map & Model-to-map & Average \\
        \rowcolor[gray]{0.95} 
        \multirow{-2}{*}{\cellcolor[gray]{0.95}Dataset} &
        (sequence similarity) & resolution (\AA) & resolution (\AA) & Q-score\\
		\hline
        \multirow{3}{*}{HA trimer}
		& 5XL8 (67\%) & 3.35 & 4.47 & 0.395 \\
		& 6IDD (47\%) & 3.27 & 3.90 & 0.435 \\
        & 2RFU (17\%) & 4.35 & 3.35 & 0.388 \\
		\hline
        \multirow{3}{*}{Na\textsubscript{X}}
		& 7XM9 (58\%) & 3.37 & 7.22 & 0.301 \\
		& 8FHD (56\%) & 3.33 & 7.22 & 0.325 \\
        & 6AGF (56\%) & 3.34 & 7.22 & 0.321 \\
		\hline
	\end{tabular}
	\caption{\textbf{The performance of cryoPROS using different initial homologous proteins.}}
	\label{table:diff_homo}
\end{table}

\subsection{Effect of the initial volume's lowpass filter cutoff frequency.}

The latent volume in cryoPROS is initialized by the lowpass filtering homologous protein, and we test the robustness of cryoPROS by varying the cutoff frequency (Nyquist, 10\AA, 20\AA, 30\AA) of the lowpass filter using two homologous proteins (PDB ID: 2RFU and 6IDD) in the HA Trimer dataset (EMPAIR-10096). The atomic model of the target protein solved from the data (EMPAIR-10097), via tilt-collection strategy is listed in PDB as 3WHE.

The sequence identity between the target (PDB ID: 3WHE) and the homolog (PDB ID: 2RFU) stands at 17\%, while between the target (PDB ID: 3WHE) and another homolog (PDB ID: 6IDD), it's 47\%.

\begin{figure}[h]
\centering 
\includegraphics[width=0.9\textwidth]{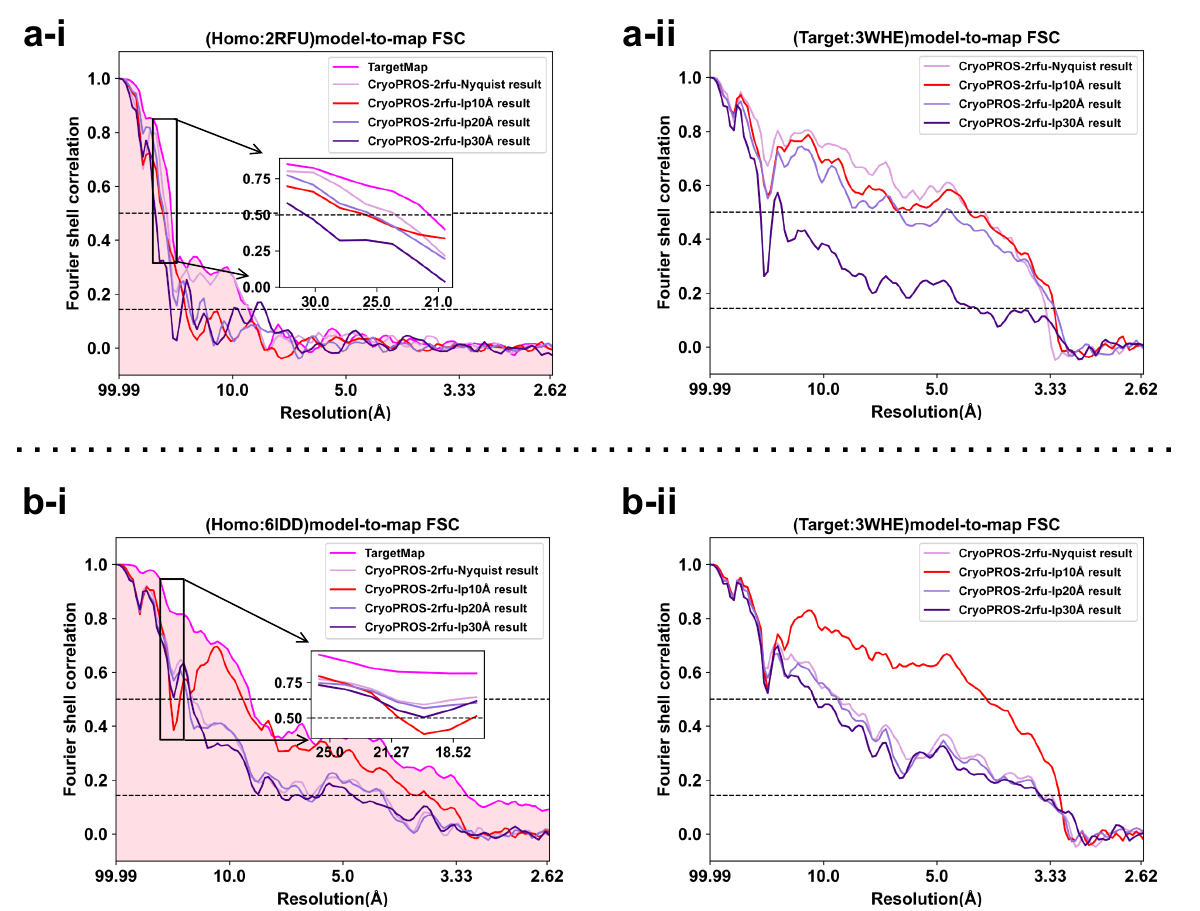}
\caption{\textbf{
Evaluating CryoPROS performance using different lowpass filter cutoff frequencies for the initial volume.} \textbf{a-i}, When employing homolog 2RFU as the initial map, we lowpass filtered the resulting map from CryoSPARC using various cutoff frequencies: Nyquist, 10\AA, 20\AA, and 30\AA. These configurations are labeled from \textsf{cryoPROS-2RFU-Nyquist} to \textsf{cryoPROS-2RFU-LP30}. For spectral similarity assessment relative to the homologous protein, we analyzed the cryoPROS-derived maps, in conjunction with the target map, through the model-to-map FSC based on the 2RFU model. The region close to the 0.5 threshold of the FSC curves is magnified and emphasized within the box. \textbf{b-i}, Analogous to \textbf{a-i}, this illustrates the spectral similarity assessment with respect to the target protein (3WHE) rather than the homologous protein (2RFU). \textbf{b}, \textbf{b} mirrors \textbf{a} (covering both \textbf{i} and \textbf{ii}), with the distinction being that 6IDD serves as the initial model in place of 2RFU.}
\label{sup_fig:lowpass}
\end{figure}

Using homolog 2RFU as the initial map, the resulting map from CryoSPARC was lowpass filtered at different cutoff frequencies, including Nyquist, 10\AA, 20\AA, and 30\AA. These are designated as \textsf{cryoPROS-2RFU-Nyquist} up to \textsf{cryoPROS-2RFU-LP30}. To assess the spectral similarity to the homologous protein, the cryoPROS-generated maps, along with the target map, were evaluated using the model (model 2RFU)model-to-map FSC, as depicted in \cref{sup_fig:lowpass}a-i. With the exception of \textsf{cryoPROS-2RFU-Nyquist}, all other FSC curves intersect the 0.5 thresholds at frequencies below 25\AA. This suggests that cryoPROS doesn't gather information from the initial volume at frequencies exceeding 25\AA, if a 10\AA or lower lowpass cutoff frequency is taken. Thus, we conclude that the model bias has not been introduced in cryoPROS.

To assess the spectral similarity to the target protein, the cryoPROS-generated maps were evaluated using the model (3WHE)-to-map FSC, as shown in \cref{sup_fig:lowpass}a-ii. While the cutoff frequency of 20\AA slightly underperforms compared to 10\AA, the performance at 30\AA is significantly worse. Based on these findings, we conclude that for this dataset, a lowpass filter cutoff frequency of 10\AA is the most suitable choice.

Meanwhile, the homolog 6IDD has a sequence similarity of 47\% with the target. Using 6IDD as the initial volume, parallel experiments reaffirm our findings (refer to \cref{sup_fig:lowpass}iia-b), strengthening the reliability of our observations. Consistent with our earlier results, a lowpass filter cutoff frequency of 10\AA\ effectively prevents information with a spectral frequency higher than 21\AA\ from influencing the cryoPROS results. Interestingly, \textsf{cryoPROS-6IDD-Nyquist} does not resemble the target protein. This suggests that for a homolog with a high sequence similarity to the target, it is prudent to apply a lowpass filter before using it as the initial volume.

Undoubtedly, the choice of this parameter warrants exploration by users across different datasets. Our observations indicate that while various lowpass filter parameters in CryoPROS don't capture distinct characteristics of homologous proteins, there are discernible differences in analytical precision. However, when balancing the need for low model bias and high analytical accuracy, a 10\AA\xspace lowpass filter for the cryoPROS initial latent volume emerges as optimal. As highlighted in the main text, a consistent 10\AA-low pass setting was adopted for all datasets. The rationale for this choice has been validated in previous sections, where the reliability of CryoPROS results was confirmed under this specific configuration.

\section{Model bias-related experiments}

\subsection{\bf \noindent Computation of KL divergence and SNR.}
To demonstrate the superiority of cryoPROS-generated noisy particles over those generated by the Gaussian noise substitution method, we compute the KL divergence between the generated and real particles and the particle SNR. Let $\{\vx^{g}_i\}_{i=1}^N$ and $\{\vx^{r}_i\}_{i=1}^N$ represent the sets of generated and real particles, respectively, with $N$ denoting the number of particles. To compute the KL divergence, we first convert the particle images to probability histograms, denoting the number of bins as $B$, and for a particle image $\vx$, its probability histogram $\vp \in \RR^{B}$. Let the probability histograms for $\{\vx^{g}_i\}_{i=1}^N$ and $\{\vx^{r}_i\}_{i=1}^N$ be represented as $\{\vp^{g}_i\}_{i=1}^N$ and $\{\vp^{r}_i\}_{i=1}^N$, respectively. Then, the average KL divergence for all particles is defined as:
\begin{equation}
    \KL(\{\vx^{g}_i\}_{i=1}^N, \{\vx^{r}_i\}_{i=1}^N) = \frac{1}{N} \sum_{i=1}^N \sum_{j=1}^{B} \vp^{g}_{ij} \log \frac{\vp^{g}_{ij}}{\vp^{r}_{ij}}
\end{equation}
where $\vp^{g}_{ij}$ and $\vp^{r}_{ij}$ denotes the $j$-th component of $\vp^{g}_i$ and $\vp^{r}_i$, respectively. In practice, since the pixel values within a particle image remain unbounded, we first clip the pixel values within the range of $[-4, 4]$, and the number of bins $B$ is set to 1024.

We estimate the SNR of noisy particles using the method proposed in~\cite{topazdenoise}. First, we randomly sample $N$ noisy particles, denoted as $\{\vx_i\}_{i=1}^N$. Subsequently, we randomly sample $N$ background particles from the original micrographs, denoted as $\{\vx^b_i\}_{i=1}^N$, which contain only pure noise. Denote the mean and variance for each background particle $\vx^b_i$ as $\vmu^b_i$ and $\vv^b_i$, respectively. Then, we normalize the noise in $\vx_i$ and convert it to $\hat{\vx}$ through the equation: $\hat{\vx}_i = \vx_i - \vmu^b_i$. We denote the mean and variance for $\hat{\vx}_i$ as $\vmu_i$ and $\vv_i$. The average SNR (dB) for the noisy particles is then defined as:
\begin{equation*}
    \mathrm{SNR}=\frac{10}{N} \sum_{i=1}^N \log_{10}\left(\vv_i\right)-\log_{10}\left(\vv^b_i\right).
\end{equation*}
In practice, we set the value of $N$ to 10.


\subsection{Gaussian noise substitution method.}

In this section, we aim to establish the necessity of employing Conditional Variational Autoencoders (CVAE) by comparing it with an alternative approach: Gaussian noise substitution method. The Gaussian noise substitution method involves substituting the learned noise in cryoPROS with Gaussian noise while keeping all other parameters unchanged. The step-by-step process of this alternative method is as follows: (1) Initialization: The lowpass filtered homologous protein serves as the initial latent volume for projection, utilizing a uniform pose direction, where the projection process was carried out using the \texttt{relion\_project} module within Relion. (2) Modulation and noise addition: CTF modulation is applied to the projected images, followed by the addition of Gaussian noise (standard deviation of added white Gaussian noise is 50, resulting in a signal-to-noise ratio (SNR) of $-14.84$dB). (3) Integration into refinement module: Gaussian noisy articles are imported into cryoPROS's refinement model, and the same processing steps are repeated. Subsequently, the latent volume is updated, initiating an iterative cycle of the aforementioned steps. We evaluate these two methods on HA-trimer and Na\textsubscript{X} datasets, and the results are shown in Table~\ref{table:gaussiannoise}. From the table, we find the Gaussian noise substitution method failed to achieve the desired quality on the Na\textsubscript{X} dataset, and for the HA trimer dataset, the Gaussian noise substitution method exhibits some potential for success, whereas the reconstructed density displayed a closer resemblance to the homologous protein than the result produced by cryoPROS, which raises concerns about the potential model bias problem.

In summary, from the results of this series of experiments, the Gaussian noise substitution method proves to be inadequate for mitigating model bias and generating accurate protein reconstructions and shows the indispensable role played by the cryoPROS generative module in averting model bias effectively and producing reliable outcomes in cryo-EM studies.

Our investigation involves conducting experiments on two distinct datasets, namely the HA-trimer and Na\textsubscript{X}. The obtained results unveil intriguing insights. The Gaussian noise substitution method does not yield satisfactory outcomes for the Na\textsubscript{X} protein (see Table~\ref{table:gaussiannoise}). The resulting reconstructions fail to achieve the desired quality, indicating the ineffectiveness of the Gaussian noise substitution method in this context. On the other hand, in the case of the HA trimer, the Gaussian noise substitution method exhibits some potential for success. However, closer examination reveals that the outcomes bear a striking resemblance to the homologous protein, rather than the results produced by cryoPROS (see Table~\ref{table:gaussiannoise}). This phenomenon indicates a notable risk of introducing model bias into the reconstructions through the Gaussian noise substitution method.

In summary, these series of experiments convincingly underscore the significance of the generative module within the cryoPROS framework. The Gaussian noise substitution method proves to be inadequate for mitigating model bias and generating accurate protein reconstructions. The results emphasize the indispensable role played by the cryoPROS generative module in averting model bias effectively and producing reliable outcomes in cryo-electron microscopy studies.

\begin{table}[]\setlength{\tabcolsep}{5pt}
\renewcommand{\arraystretch}{1.5}
\centering
\begin{tabular}{ccccc}
\hline
\rowcolor[gray]{0.95}
& \multicolumn{2}{c}{HA trimer} & \multicolumn{2}{c}{Na\textsubscript{X}}       \\
\rowcolor[gray]{0.95}
                        & Target model  & Homolog model & Target model & Homolog model  \\
\rowcolor[gray]{0.95}
\multirow{-3}{*}{Method} & (3WHE) & (6IDD) & (Na\textsubscript{X}) & (Nav1.6: 8FHD) \\ \hline
GN substitution & 4.14 & 9.58 & 40.23 & 31.28 \\
CryoPROS & 4.14 & 20.96 & 7.22 & 23.47 \\ \hline
\end{tabular}
\caption{\textbf{Comparison of model-to-map resolutions (\AA) between density maps reconstructed using Gaussian noise substitution method and cryoPROS. Higher resolution is preferred for the target model, while lower resolution is preferred for the homolog model.}}
\label{table:gaussiannoise}
\end{table}


\subsection{Model bias verification of other datasets.}

In main text, we employed a structural refinement method to validate the accuracy of cryoPROS resulting map in the case of  HA trimer. This validation demonstrated a remarkable closeness between the cryoPROS outcomes and the true structure, while simultaneously showcasing a significant dissimilarity from homologous proteins structure. This evidence substantiates the claim that cryoPROS is devoid of model bias risks.

\begin{figure}[h]
\centering 
\includegraphics[width=0.9\textwidth]{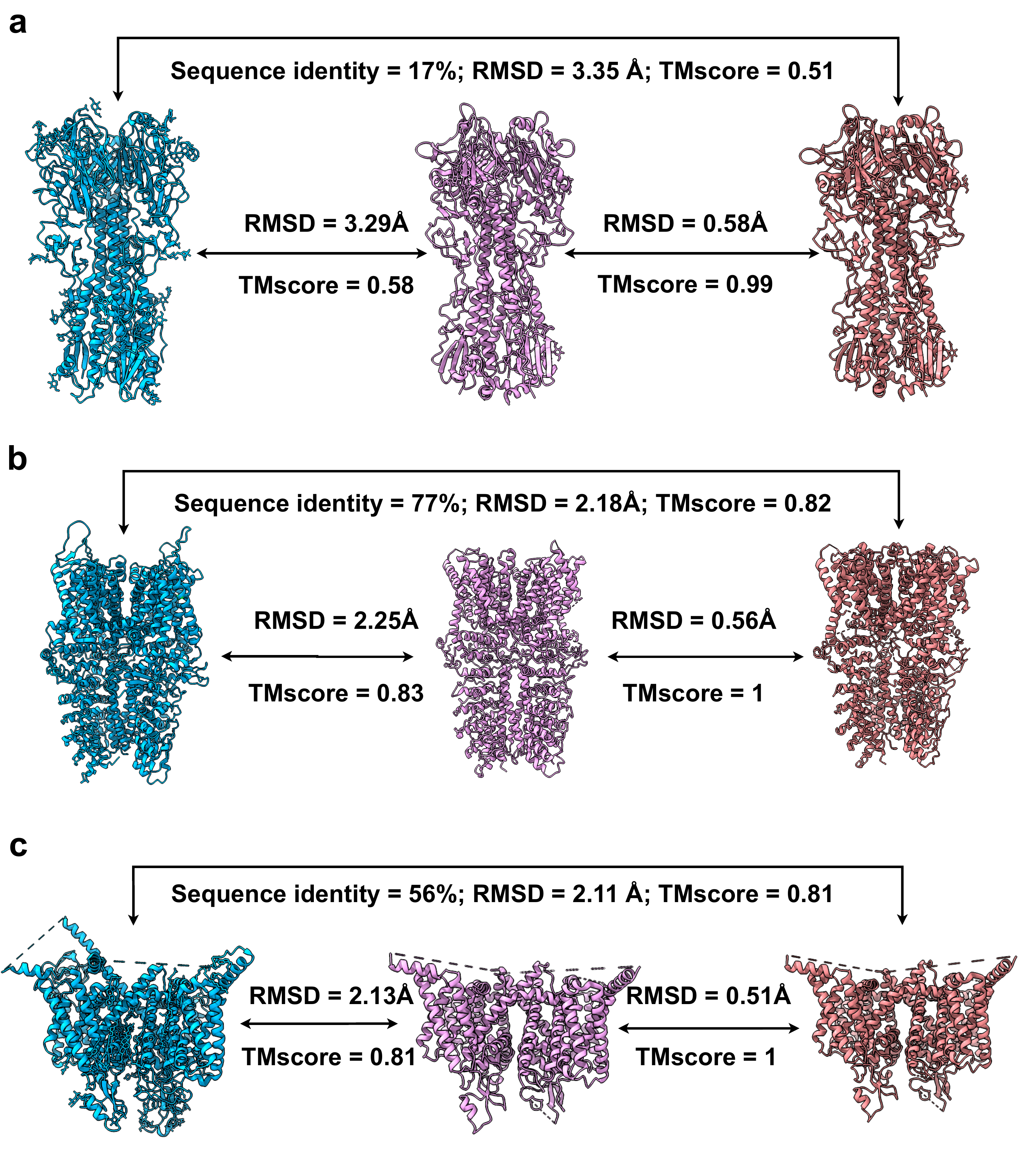}
\caption{\textbf{Structure refinement experiments on other datasets.} \textbf{a}, For the SIM2 dataset, a comparative analysis is presented involving the atomic models of the homologous protein, a structure-refined model derived from the cryoPROS resulting map (based on target model), and the target protein. The similarity between the two models is evaluated through Root Mean Square Deviation (RMSD) and TMscore calculations. \textbf{b,c}, Similar results are observed for the PO-subset of EM-10024 and Na\textsubscript{X} datasets.}
\label{sup_fig:modelbias}
\end{figure}

Expanding upon this notion, we extended our investigation to encompass three additional datasets: SIM2 (see \cref{sup_fig:modelbias}a), the PO-subset of EM-10024 (see \cref{sup_fig:modelbias}b), and Na\textsubscript{X} (see \cref{sup_fig:modelbias}c).  We found that the experiment had consistent results across different data sets. These results further reinforce our earlier assertion about cryoPROS's proficiency in generating unbiased results.  It reinforces its capability to produce accurate reconstructions that remain faithful to the target structures, while mitigating the risks associated with model bias.

\section{Hierarchical structure of CVAE}
\label{sec:hs}
\begin{figure}
\centering 
\includegraphics[width=1\textwidth]{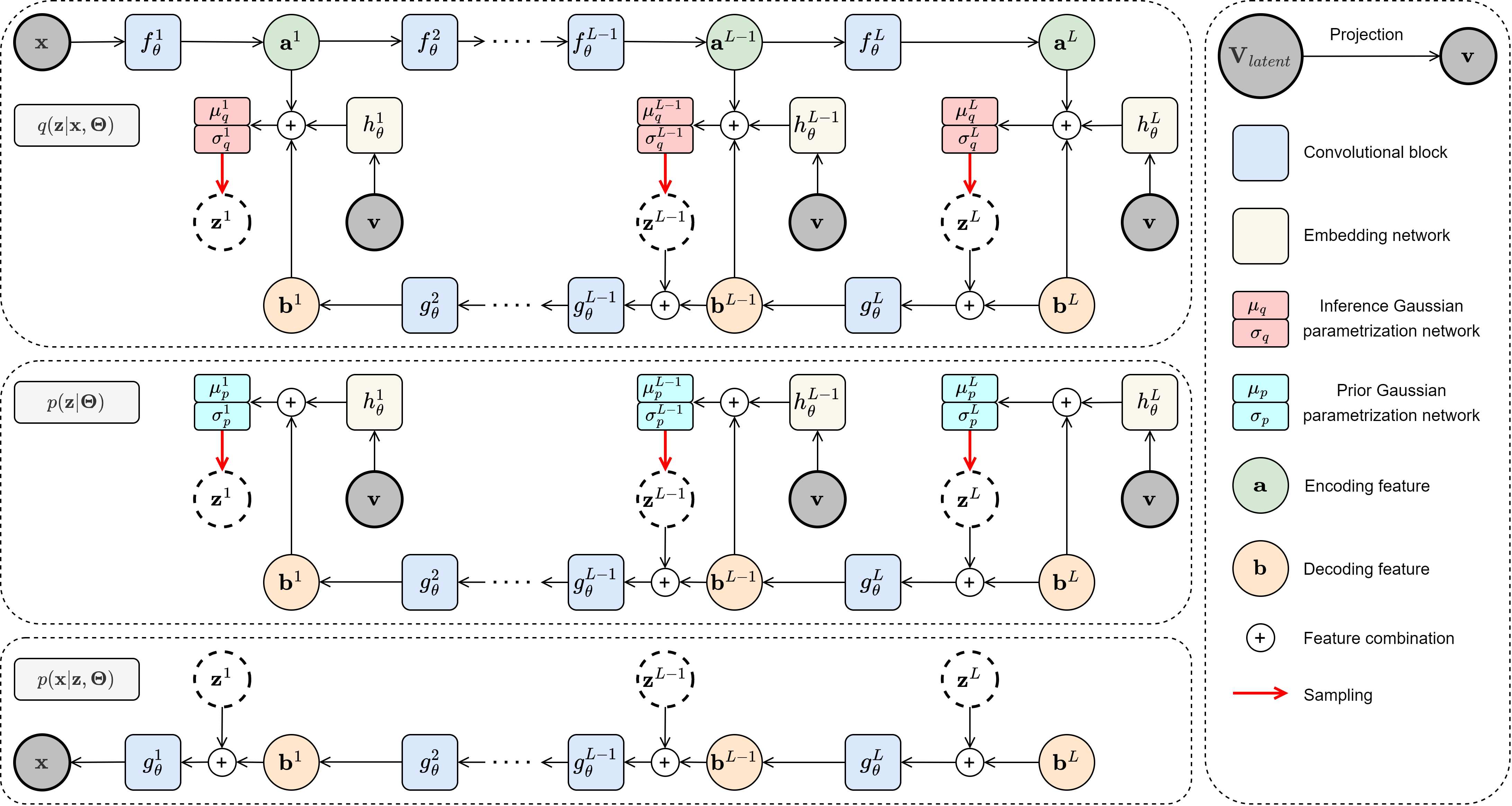}
\caption{\textbf{Network architecture of the hierarchical CVAE model, including the inference model $q(\vz\lvert\vx,\vTheta)$, the prior model $p(\vx\lvert \vTheta)$, and the generative model $p(\vx\lvert\vz,\vTheta)$.}}
\label{hvae_arch}
\end{figure}

In cryoPROS, a hierarchical structure~\cite{NEURIPS2020_e3b21256,child2021very} is adopted for the CVAE consisting consists of $L$ stochastic layers. Specifically, we represent the latent variable as $\vz = (\vz^1,\dots,\vz^L)$, and the prior model is decomposed as follows:
\begin{equation}
	p(\vz\lvert\vTheta) = p(\vz^L\lvert\vTheta) \prod_{l=1}^{L-1} p(\vz^{l}\lvert\vz^{>l}, \vTheta)
\end{equation}
where $\vz^{>l} = (\vz^{l+1}, \dots,\vz^{L})$. Similarly, the inference models follow the same order:
\begin{equation}
	q(\vz\lvert\vx,\vTheta) = q(\vz^L\lvert\vx,\vTheta) \prod_{l=1}^{L-1} q(\vz^{l}\lvert\vz^{>l}, \vx, \vTheta).
\end{equation}
Then, the cELBO can be expressed as:
\begin{equation}
	\begin{aligned}
		\text{cELBO} =& \EE_{q(\vz\lvert\vx, \vTheta)} \log p(\vx \lvert \vz, \vTheta) - \KL(q(\vz^L\lvert\vx, \vTheta)\|p(\vz^L\lvert\vTheta)) \\
		&- \sum_{l=1}^{L-1} \EE_{q(\vz^{>l} \lvert \vx, \vTheta)} \KL(q(\vz^{l}\lvert \vz^{>l}, \vx, \vTheta) \| p(\vz^{l}\lvert \vz^{>l}, \vTheta)).
	\end{aligned}
\end{equation}
Regarding the inference model $q(\vz\lvert\vx,\vTheta)$, we assume the following form:
\begin{equation}
	q(\vz^{l}\lvert\vz^{>l}, \vx, \vTheta) = \cN(\mu^l_q(\va^l, \vb^l, h_\theta^l(\vv)), \sigma^l_q(\va^l, \vb^l, h_\theta^l(\vv))), \quad l=1,2,\dots, L
\end{equation}
where $\va^l$ and $\vb^l$ denotes the encoding and decoding feature in $l$-th layer, respectively. $\vv=\vC(\psi) P(\phi) \vV_{latent}$ denotes the projection of the latent volume, and it is embedded into the latent space using a neural network denoted as $h_\theta(\cdot)$. $\mu^l_{q}$ and $\sigma^l_{q}$ are networks that convert $(\va^l, \vb^l, h_\theta^l(\vv))$ to the parameters of a Gaussian distribution. The encoding features $\{\va^l\}_{l=1}^L$ are recursively obtained as follows:
\begin{equation}
	\va^{1} = f_{\theta}^{1}(\vx), \quad \va^{l} = f_{\theta}^{l}(\va^{l-1}), \quad l=2,\dots,L,
\end{equation}
where $f_\theta^l$ represents the convolutional block in the $l$-th encoding layer. The decoding features $\vb^l$ are obtained through the recursion:
\begin{equation}
	\vb^{l-1} = g_{\theta}^l(\vz^{l}, \vb^l), \quad l = 2,\dots,L,
\end{equation}
where $\vz^{l}$ is sampled from $\cN(\mu_{q}^l(\va^l, \vb^l, h_\theta^l(\vv)), \sigma_{q}^l(\va^l, \vb^l), h_\theta^l(\vv))$, $\vb^{L}$ is a constant vector that is set as a learnable parameter, and $g_{\theta}^l$ is the convolutional block in $l$-th decoding layer. Additionally, for the prior model $p(\vz\lvert\vTheta)$, we assume the following form:
\begin{equation}
	p(\vz^l\lvert \vz^{>l}, \vTheta) = \cN(\mu^l_p(\vb^l, h_\theta^l(\vv)), \sigma^l_p(\vb^l, h_\theta^l(\vv))), \quad l=1,2,\dots,L,
\end{equation}
and the generative model $p(\vx\lvert\vz,\vTheta)$ is assumed to be $\cN(g_\theta^1(\vz^1,\vb^1), \vI)$. We adopt the Residual Dense Block (RDB)~\cite{8578360} as our convolutional block for $f_{\theta}^l$ and $g_\theta^l$, and the model architecture is shown in \cref{hvae_arch}. The loss function for the CVAE aims to minimize the $-\text{cELBO}$, and with the given parametrization, the loss function is:
\begin{equation}
    \text{Loss}(\vx,\vTheta) = \EE_{q(\vz\lvert\vx,\vTheta)} \| g_{\theta}^1(\vz^1,\vb^1) - \vx \|_2^2 + \sum_{l=1}^{L} \KL \left(\cN(\mu_{q}^l, \sigma_{q}^l) \| \cN(\mu_{p}^l, \sigma_{p}^l) \right).
\end{equation}
The KL divergence for two Gaussian distributions has an analytical form, which is:
\begin{equation}
\begin{aligned}
    \KL \left(\cN(\mu_{q}, \sigma_{q}) \| \cN(\mu_{p}, \sigma_{p}) \right) =& \frac{1}{2} \left( \log (\lvert\sigma_p\rvert / \lvert\sigma_q\rvert) - K \right. \\
    & \left.+ \mathrm{Tr}(\sigma_p^{-1}\sigma_q) + (\mu_q - \mu_p)^{T}\sigma_p^{-1}(\mu_q - \mu_p) \right),
\end{aligned}
\end{equation}
where $K$ denotes the dimension of the latent variable $\vz$.

\bibliography{2_supplementary.bib}